\documentclass[apj]{emulateapj}
\usepackage{amsmath}
\usepackage{amssymb}
\usepackage{graphicx}

\usepackage{graphicx}
\usepackage{natbib}
\def\aap{A\&A} 
\def\aj{AJ} 
\def\apj{ApJ} 
\def\apjl{ApJL} 
\def\apjs{ApJS} 
\def\mnras{MNRAS} 
\def\nat{Nature} 

\def\K{{\rm K}} 
\def\hr{{\rm hr}}
\def\Ms{{\rm M}\s} 
\def\GHz{{\rm GHz}} 
\def\m{{\rm m}} 
\def\mm{{\rm m}\m} 
\def\cm{{\rm c}\m} 
\def\km{{\rm k}\m} 
\def\pc{{\rm pc}} 
\def\Mpc{{\rm M}\pc} 
\def\Ms{M_\odot} 
\def\mas{{\rm mas}} 
\def\muas{\mu{\rm as}} 
\def\rad{{\rm rad}}

\newcommand\bmath[1] {\mbox{\boldmath$\rm #1$}}
\def\d{{\rm d}}
\def\e{{\rm e}}

\def\j{\bmath{j}}
\def\v{\bmath{v}}
\def\E{\bmath{E}}
\def\B{\bmath{B}}

\def\div{\bmath{\nabla}\cdot}
\def\grad{\bmath{\nabla}}
\def\curl{\bmath{\nabla}\times}

\def\ph{\bmath{\phi}}

\def\em{\it}

\def\Fs{{}^*\!F}

\begin{document}

\title{Imaging the Black Hole Silhouette of M87:\\ Implications for Jet
Formation and Black Hole Spin}

\author{
Avery E. Broderick\altaffilmark{1} \&
Abraham Loeb\altaffilmark{2}
}

\altaffiltext{1}{Canadian Institute for Theoretical Astrophysics, 60 St.~George St., Toronto, ON M5S 3H8, Canada; aeb@cita.utoronto.ca}
\altaffiltext{2}{Institute for Theory and Computation, Harvard University, Center for Astrophysics, 60 Garden St., Cambridge, MA 02138.}

\shorttitle{Imaging the Black Hole Silhouette of M87}
\shortauthors{Broderick \& Loeb}

\begin{abstract}

The silhouette cast by the horizon of the supermassive black hole in
M87 can now be resolved with the emerging millimeter very-long baseline
interferometry (VLBI) capability.  Despite being $\sim 2\times 10^3$
times farther away than Sgr A* (the supermassive black hole at the
center of the Milky-Way and the primary target for horizon-scale
imaging), M87's much larger black hole mass results in a horizon
angular scale roughly half that of Sgr A*'s, providing another practical
target for direct imaging.  However, unlike Sgr A*, M87 exhibits a powerful
radio jet, providing an opportunity to study jet
formation physics on horizon scales.  We employ a simple,
qualitatively correct force-free jet model to explore the expected
high-resolution images of M87 at wavelengths of $1.3\,\mm$ and
$0.87\,\mm$ ($230\,\GHz$ and $345\,\GHz$), for a variety of jet
parameters.  We show that future
VLBI data will be able to constrain the size of the jet footprint, the
jet collimation rate, and the black hole spin.  Polarization will
further probe the structure of the jet's magnetic field and its effect
on the emitting gas.  Horizon-scale imaging of M87 and Sgr A* will
enable for the first time the empirical exploration of the
relationship between the mass and spin of a black hole and the
characteristics of the gas inflow/outflow around it.

\end{abstract}

\keywords{black hole physics --- techniques: interferometric --- submillimeter --- galaxies: active --- galaxies: jets}

\maketitle

\section{Introduction}\label{sec:intro}
With the advent of millimeter VLBI, for the first time it has become
possible to resolve the horizon of a black hole
\citep{Doel_etal:08,Doel:08}.
Most of the theoretical and observational efforts on this front have
been directed towards imaging Sagittarius A* (Sgr A*), the radio
source associated with the supermassive black hole at the center of
the Milky Way
\citep[see, e.g.,][]{Falc-Meli-Agol:00,Brod-Loeb:05,Brod-Loeb:06a,Brod-Loeb:06b,Miyo_etal:08,Doel_etal:08}.
This is because Sgr A* subtends the largest angle of any known black
hole candidate (roughly $55\,\muas$ in diameter).  Nevertheless, there
is at least one other promising target for millimeter VLBI imaging,
namely M87's supermassive black hole (hereafter labeled `M87' for
brevity).

At a distance of $16\,\Mpc$ and with a black hole mass of
$3.4\times10^9\,\Ms$, the apparent diameter of M87's horizon is
approximately $22\,\muas$, appearing about half as large as Sgr A*.
Like Sgr A*, M87 is many orders of magnitude under-luminous in
comparison to its Eddington luminosity.  However, M87 is different
from Sgr A* in five important respects.

First, M87 is located in the
Northern sky, and thus is more amenable to imaging by telescope arrays
comprised of existing millimeter and submillimeter observatories,
which exist primarily in the Northern hemisphere.  Thus, observations
of M87 will generally have lower atmospheric columns and longer
observing opportunities.  Arrays of existing telescopes (see
\S\ref{sec:PT} for details), spanning projected baselines of
$10^4\,\km$, are already sufficient to resolve the silhouette at
$1.3\,\mm$ ($230\,\GHz$) and $0.87\,\mm$ ($345\,\GHz$) with beam sizes
of $17\,\muas$
and $11\,\muas$, respectively, along the minor axis.  The inclusion of
new facilities over the next few years will improve baseline
coverage substantially, resulting in a roughly circular beam with
these resolutions.

Second, M87's mass implies a dynamical timescale of roughly
$5\,\hr$, and a typical orbital timescale at the innermost-stable
orbit of 2-18 days, depending upon the black hole spin.  Thus, unlike
Sgr A*, the structure of M87 may be treated as fixed during an entire
day, one of the underlying assumptions of Earth aperture synthesis.
This means that it will be possible to produce sequences of images of
dynamical events in M87, occuring over many days.

Third, M87 exhibits a powerful jet, providing a unique opportunity
to study the formation of relativistic jets on horizon scales.
Realistic {\em ab initio} computations of jet formation are not
presently possible, requiring jet simulations to make a variety of
assumptions about the underlying plasma processes.  Examples include
assumptions about the dissipation scale, cooling rate, jet
mass-loading mechanism, acceleration of the observed non-thermal
electrons and even the applicability of the magnetohydrodynamic (MHD)
prescription
\citep{DeVi-Hawl-Krol:03,McKi-Gamm:04,Komi:05,McKi:06,Hawl-Krol:06,Tche-McKi-Nara:08,Igum:08,McKi-Blan:08,Nish_etal:05}. 
Thus, horizon-resolving images of M87's jet will critically inform our
understanding of the formation, structure and dynamics of
ultra-relativistic outflows.

Fourth, unlike Sgr A*, one is primarily driven towards submillimeter
wavelengths by resolution requirements.  There is no evidence for an
analogous interstellar scattering screen that would blur high
resolution images  of M87 \citep[cf.][]{Bowe_etal:06}.
Nor is there evidence that M87 is optically thick at or about
millimeter wavelengths \citep[cf.][]{Brod-Loeb:06a,Brod-Loeb:06b}.
Thus, modeling and interpreting images of M87 is likely to be somewhat
more straightforward.

Fifth, despite being under-luminous, M87 is similar to other
radio-loud quasars.  The same cannot be said of Sgr A*, which is
nearly dormant and is visible only due to its extraordinary proximity.
As such, the faint low-mass black hole in Sgr A* is much more
representative of the population of black holes that are accessible to
the {\em Laser Interferometer Space Antenna} (LISA
\footnote{http://lisa.nasa.gov/}).  Careful comparison between
horizon-scale imaging of Sgr A* and M87 will be crucial as a
touchstone for comparing the astrophysical properties of these two
populations, and how the black hole mass and spin affect the
electromagnetic and kinetic properties of the surrounding plasma.

It is not surprising, therefore, that already high-resolution images
of M87 has been aggressively pursued.  The best known of these are the
$7\,\mm$ VLBI images obtained by \citet{Juno-Bire-Livi:99} \citep[and
more recently by][]{Ly-Walk-Wrob:04,Walk-Ly-Juno-Hard:08}.  $3\,\mm$
and $2\,\mm$ images have also been reported
\citep{Kric_etal:06,Kova-List-Homa-Kell:07}.  These represent the
highest resolution observations of a relativistic jet ever produced, probing
down to roughly 5 Schwarzschild radii.  Nevertheless, they are
presently insufficient to directly address many outstanding questions
in jet-formation theory.  Among these questions are the role of the
black hole spin in jet production, the structure of the jet forming
region, the importance of magnetic fields and the particle content of
the jet itself.  However, all of these uncertainties may be settled by
pushing to submillimeter wavelengths.  In this sense, millimeter
imaging promises to provide a qualitative change, not simply an
incremental improvement.  The reason for this is simple: the minimum
relevant physical scale for a black hole of mass $M$ is $GM/c^2$, and
thus resolving this scale provides full access to the physical
processes responsible for jet formation and accretion.
In addition, the observation of the characteristic silhouette
associated with the black hole settle the present ambiguity in
indentifying its position relative to the observed radio core
\citep{Marc_etal:08}.

Motivated by the above arguments, we present in this paper predicted
theoretical images of M87 for a class of force-free jet models which
fit all existing observations.  In particular, we assess the ability
of possible millimeter \& submillimeter VLBI experiments to
distinguish different critical jet parameters.  In \S\ref{sec:PT} we
review how a millimeter array can be created out of
existing telescopes.  In \S\ref{sec:MM} we describe the jet-disk model
we employ for M87.  In \S\ref{sec:JI} and \S\ref{sec:JP} we present
images and polarization maps, respectively, for a variety of jet
models.  Finally, \S\ref{sec:C} summarizes our main conclusions.

\section{Potential Millimeter Arrays} \label{sec:PT}
A detailed discussion of existing telescopes that may serve as
stations in a millimeter/submillimeter Very-Long Baseline Array (VLBA)
can be found in \citet{Doel_etal:08b}. These facilities are: CARMA in
Cedar Flat, California; the SMTO on Mount Graham, Arizona; the LMT on
Sierra Negra, Mexico; APEX \& ASTE, soon to be followed by ALMA, on
the Atacama plane in Chile; the IRAM dish on Pico Veleta (PV), Spain;
the IRAM Plateau de Bure (PdB) array in France; and a number of
telescopes atop Mauna Kea, including the JCMT, SMA and CSO, that may
be used independently or phased together to provide single Hawaiian
station.  The millimeter flux from M87 is a factor of 2--3 lower than
Sgr A*, and thus telescope sensitivity will be a key factor in
resolving small-scale structure.  This may be achieved in a variety of
ways, including phasing together the individual telescopes in the
CARMA, ALMA, SMA and PdB into single apertures at their respective
locations.  In addition, a shift to large-bandwidths will likely be
required.

\begin{figure}
\begin{center}
\includegraphics[width=\columnwidth]{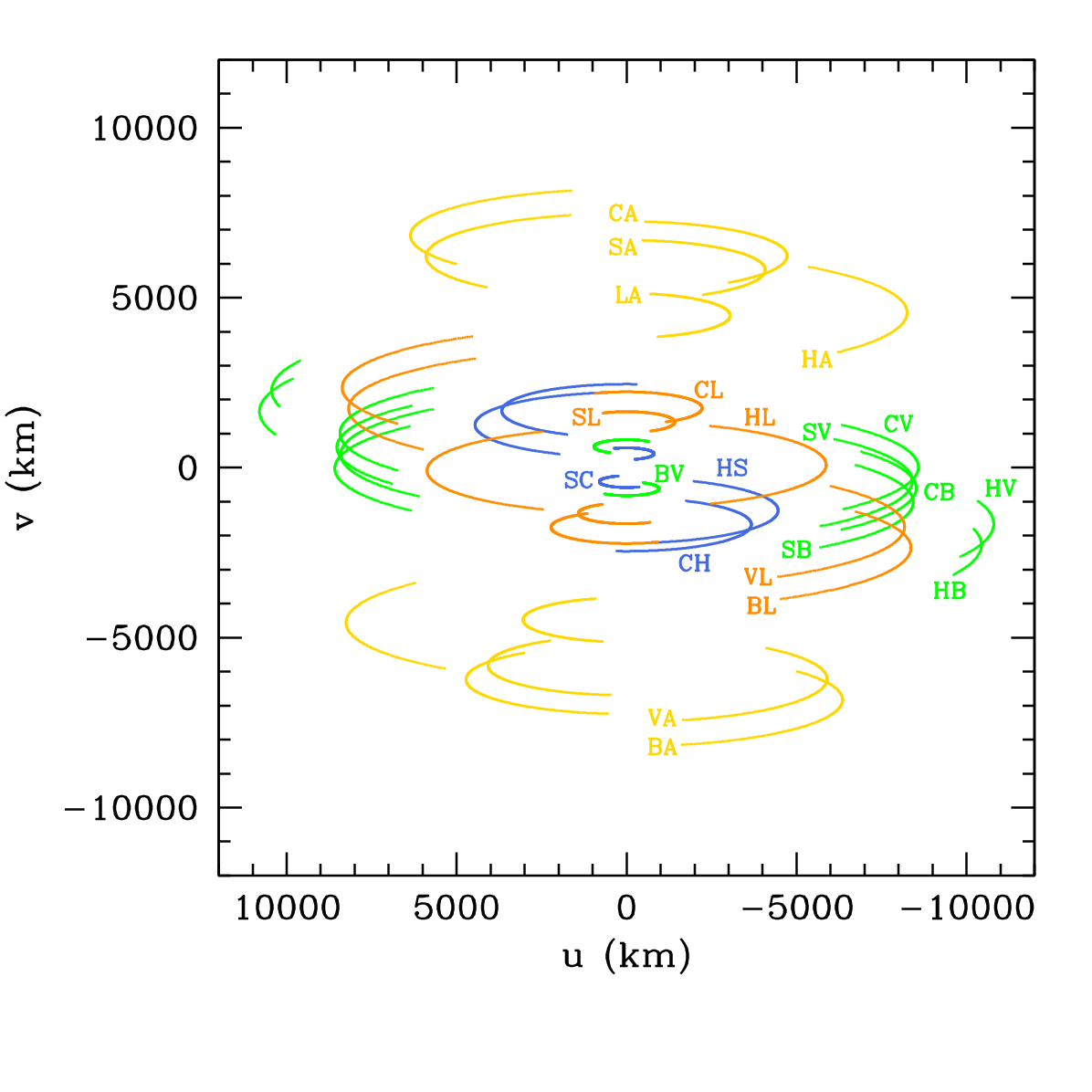}
\end{center}
\caption{Tracks in the $u$-$v$ plane (representing the
Fourier-transform coordinates of the sky) made as the Earth rotates by
existing and forthcoming telescopes as seen by M87.  Blue tracks are
associated solely with the existing North American telescopes (SMA,
JCMT, CARMA \& SMTO), green includes the two European observatories
(PdB \& PV), orange includes the LMT and yellow includes
telescopes in Chile (APEX \& ALMA).  The tracks are labelled according
to H (CSO, SMA, JCMT), C (CARMA), S (SMTO), B (PdB), V (PV), L (LMT)
and A (APEX, ASTE, ALMA).} \label{fig:uv}
\end{figure}

As mentioned in \S\ref{sec:intro}, M87 has a number of observational
advantages over Sgr A*, the primary one being its location in the
Northern sky.  This is evident in the baseline coverage provided by
the previously mentioned telescopes, shown in Fig. \ref{fig:uv}.  M87
is simultaneously visible at the European stations (PV and PdB) and
the North American stations (Hawaii, CARMA and SMTO).  With the
inclusion of the LMT, the baseline coverage is nearly complete within
a narrow, roughly East-West ellipse, providing a resolution in this
direction of approximately $17\,\muas$ and $11\,\muas$ at $1.3\,\mm$
and $0.87\,\mm$, respectively.  Introducing a station in Chile results
in a similar resolution along the North-South direction.

Critical for Earth aperture synthesis, M87's substantially larger mass
implies that it may be regarded as a stationary source, even on
horizon scales, for timescales of order a day.  The sparseness of the
baseline coverage of Sgr A*, which varies over $15\,\min$ timescales,
is due not only to the limited number of existing stations, but also
the fact that visibilities taken throughout the night cannot generally
be assumed to arise from the same intrinsic brightness distribution.
Thus, in the absence of additional telescopes, efforts to interpret
features in the image of Sgr A* \citep[such as a black hole silhouette,][]{Mill:62},
will need to be performed in Fourier space.  In contrast, M87's
comparably long dynamical timescale means that images can be produced
and analyzed directly.  This is even true during dynamical periods in
which the accretion flow is changing as rapidly as it can around the
horizon.  For example, even in the case of a maximally spinning black
hole the light-crossing time of the disk inner edge, taken to be the
Innermost Stable Circular Orbit (ISCO), is roughly $1.2\,{\rm days}$
(accounting for the strong gravitational lensing), making
large-amplitude intraday variability unlikely.

\section{Modeling M87} \label{sec:MM}

\begin{figure}
\begin{center}
\includegraphics[width=\columnwidth]{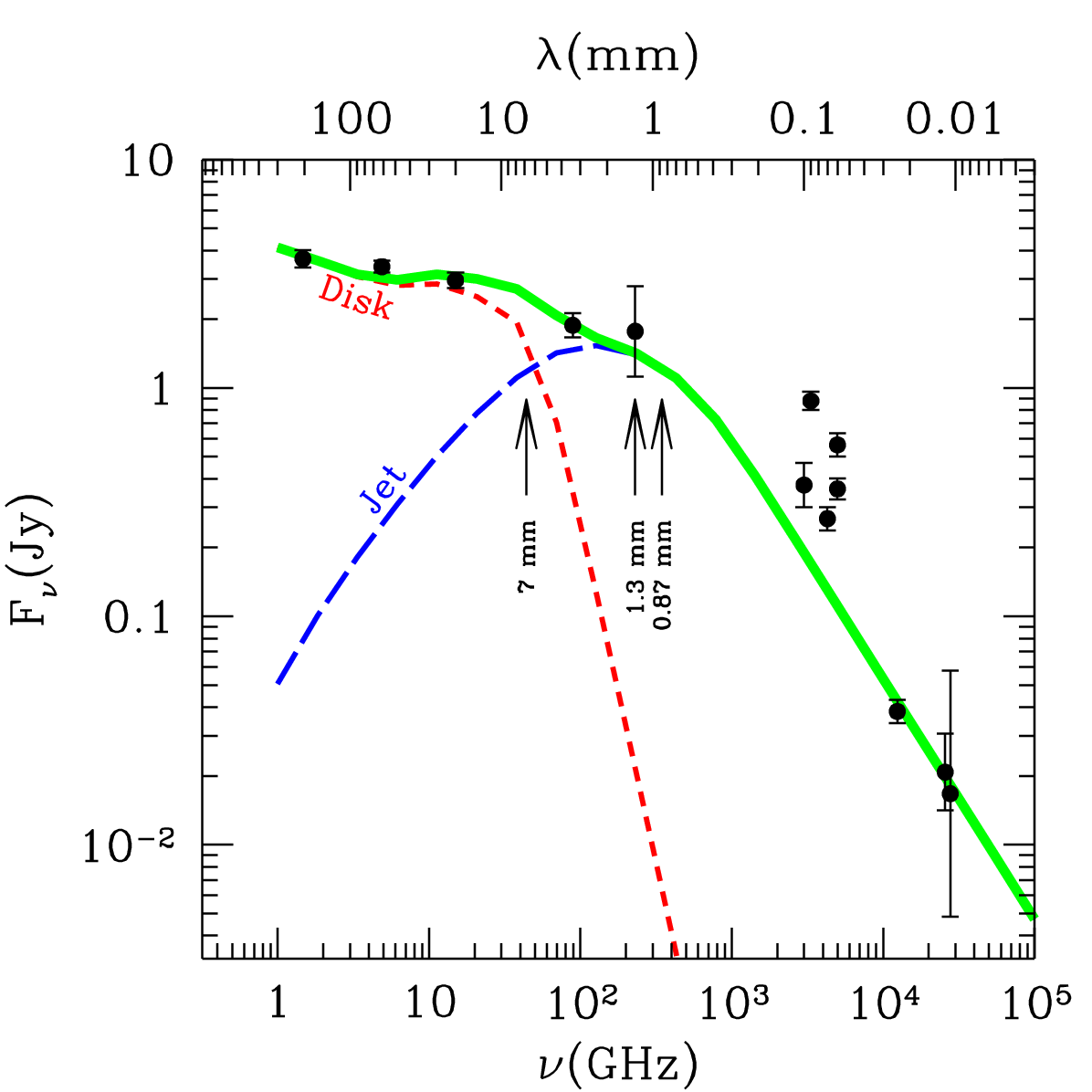}
\end{center}
\caption{Spectra of our canonical jet-disk model.  The disk \& jet components
  are shown by the short-dashed, red line and the long-dashed, blue
  line, respectively.  The total spectrum is shown by the thick green
  line.  Data points are collected from \citet{Bire-Ster-Harr:91,Desp-Frai-Davo:96,Perl_etal:01,Perl_etal:07,Tan_etal:08}. The errors on the data points
  represent the instrumental uncertainty, not the variability, and
  thus are an underestimate for the multi-epoch SED.  The spectra of the
  other jet-disk models are nearly indistinguishable from the one
  shown.} \label{fig:spec}
\end{figure}

There are a number of observational constraints upon any model for
M87.  Foremost among these is the observed Spectral Energy
Distribution (SED), shown in Fig. \ref{fig:spec}, exhibiting a nearly
flat, but not inverted, radio spectrum and a steep non-thermal decline
from millimeter to infrared wavelengths.  It is important to note that
there is no contemporaneous measurement of M87's SED from the radio to
the infrared, the points shown in Fig. \ref{fig:spec} having been
collected over nearly two decades.  The error bars shown represent the
instantaneous measurement errors, and are considerably smaller than
the observed variability.  Further difficulties arise in isolating the
``nuclear'' emission, that arising in the vicinity of the black hole,
from the emission due to knots in the parsec-scale structure of the
jet and dust in the nucleus of the surrounding galaxy.  This latter
effect is especially problematic between $24$ and $100$ microns
($3\times 10^3$--$10^4$GHz), where the SED is dominated by nuclear
dust emission \citep{Perl_etal:07}.  For this reason, we do not
attempt to fit the SED at these wavelengths.

In addition to the SED, there are a number of $7\,\mm$ VLBA images,
showing the beginnings of the jet on milliarcsecond scales
\citep{Juno-Bire-Livi:99,Ly-Walk-Wrob:04,Walk-Ly-Juno-Hard:08}.
Any viable model for M87 must be able to 
reproduce this extended $7\,\mm$ emission, and in particular showing
the wide jet morphology.

In Eddington units, M87 is significantly under-luminous for an
accreting black hole, though considerably more luminous than a
rescaled Sgr A*.  This may be in part due to the presence of a jet.
However, a number of authors have found that M87's SED is best fit by
a two-component model consisting of a Radiatively Inefficient
Accretion Flow (RIAF)-type disk producing the flat radio spectrum, and
a jet producing the millimeter, infrared and optical spectrum
\citep{Yuan:00,Reyn-diMa-Fabi:99}.  The requirement that these models
be able to qualitatively reproduce the $7\,\mm$ VLBA image implies
that the jet luminosity must be comparable to, if not dominate, that
from the disk component at this wavelength.  In the following
subsections we describe our models for the disk and jet structures,
their emission and our effort to fit all observational constraints
simultaneously.

\subsection{Radiative Transfer} \label{sec:MM.RT}
The primary emission mechanism is synchrotron, arising from both
thermal and non-thermal electrons.  We model the emission from the
thermal electrons using the emissivity described in \citet{yuan03},
appropriately altered to account for relativistic effects \citep[see,
e.g.,][]{Brod-Blan:04}.  Since we perform polarized radiative transfer
via the entire complement of Stokes parameters, we employ the
polarization fraction for thermal synchrotron as derived in
\citet{petrosian83}.  In doing so we have implicitly assumed that the
emission due to thermal electrons is isotropic, but this simplifying
assumption is not expected to affect our results significantly.

Regarding the non-thermal electrons, we follow \citet{jones77} in
assuming a power-law electron distribution with a cut-off below some
minimum Lorentz factor.  As shown below, the corresponding
low-frequency cut-off is critical to fitting M87's millimeter
spectrum.  For both thermal and non-thermal electrons, the
absorption coefficients are determined directly via Kirchoff's law
\citep{Brod-Blan:04}.

\subsection{Disk} \label{sec:MM.D} 
Given the low core luminosity of M87 in Eddington units, we employ a
radiatively-inefficient accretion flow model similar to the one we
have previously used for Sgr A* \citep{Brod-Loeb:06a,Brod_etal:08}.
This model is adequate at the low accretion rate of interest here and
makes use of the homologous nature of black hole accretion flows.
Furthermore, this model incorporates, in a limited sense, the presence
of outflows, clearly appropriate for M87.  Unlike Sgr A*, M87's radio
spectrum is not inverted, and so the model parameters are different.
It is important to emphasize that our images are largely insensitive
to the particular model we adopt for M87's accretion flow, since at
millimeter wavelengths M87 is dominated by the jet emission.  Thus,
while we include a disk model for completeness, alternative accretion
models produce nearly identical results at millimeter wavelengths.

\begin{figure*}
\begin{center}
\includegraphics[width=0.48\textwidth]{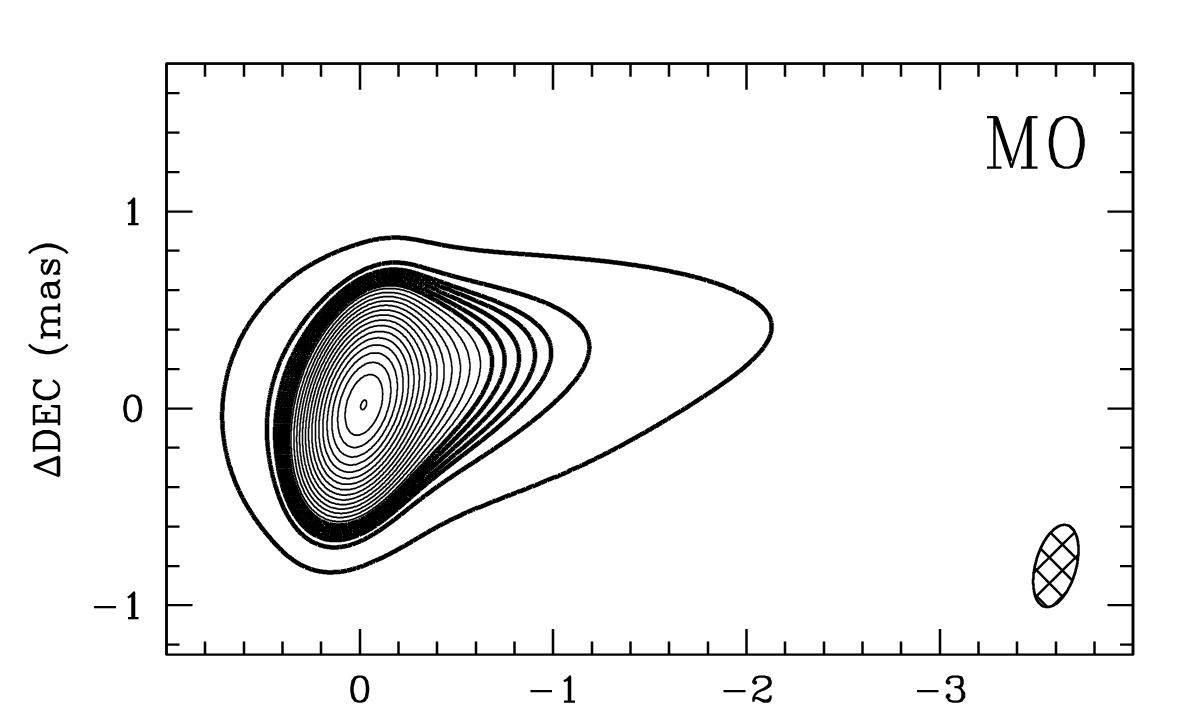}
\includegraphics[width=0.48\textwidth]{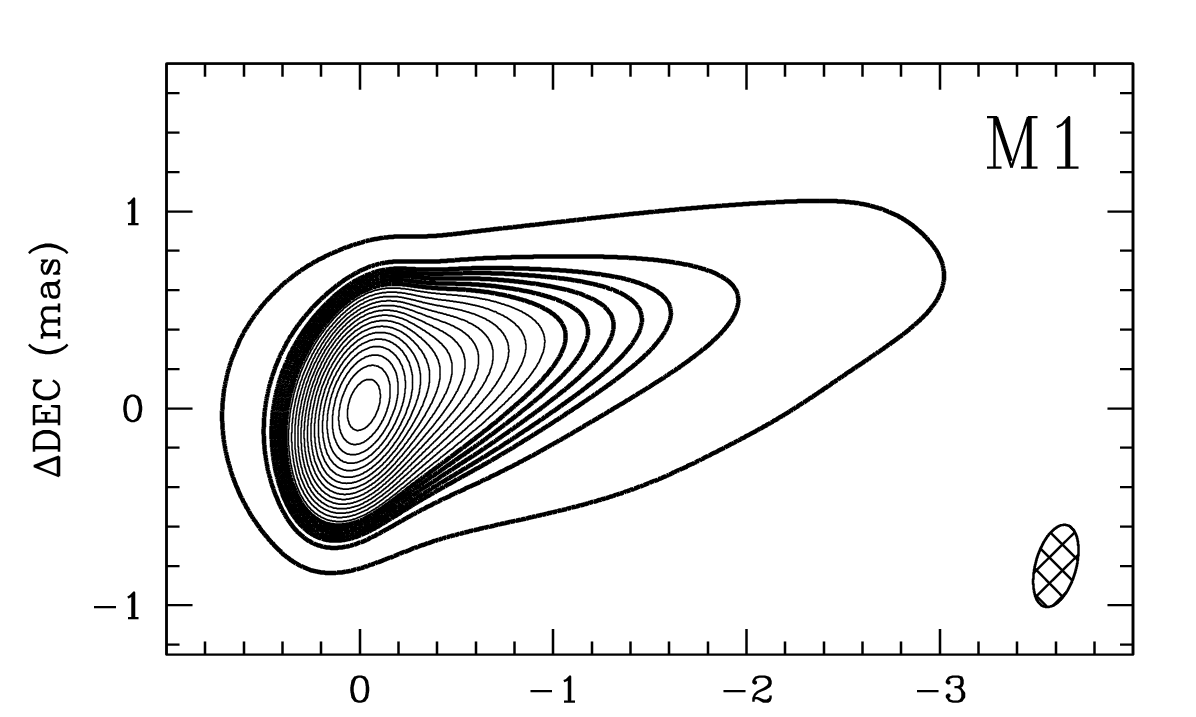}\\
\includegraphics[width=0.48\textwidth]{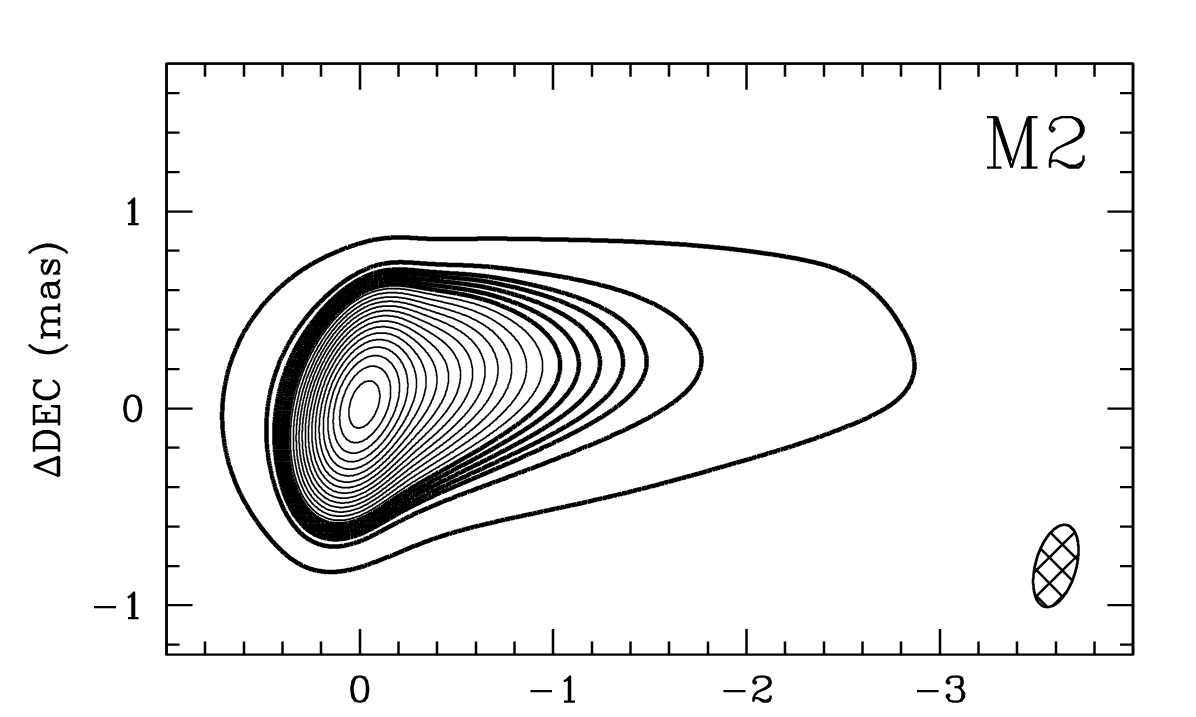}
\includegraphics[width=0.48\textwidth]{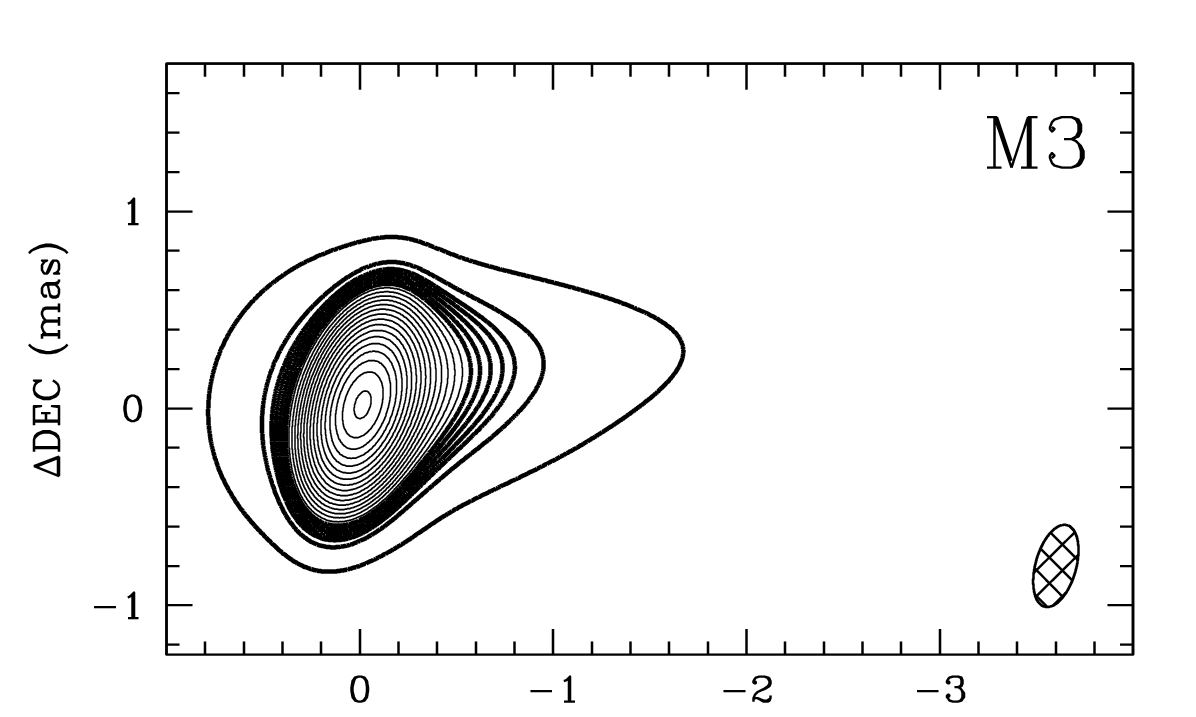}\\
\includegraphics[width=0.48\textwidth]{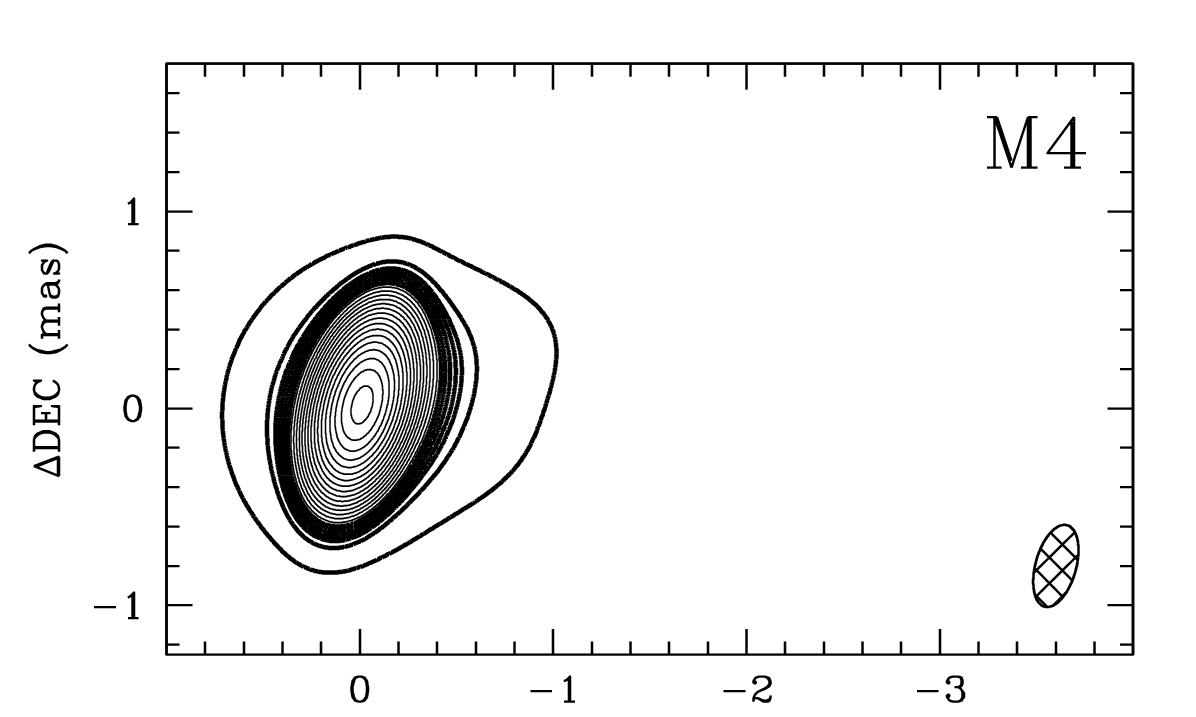}
\includegraphics[width=0.48\textwidth]{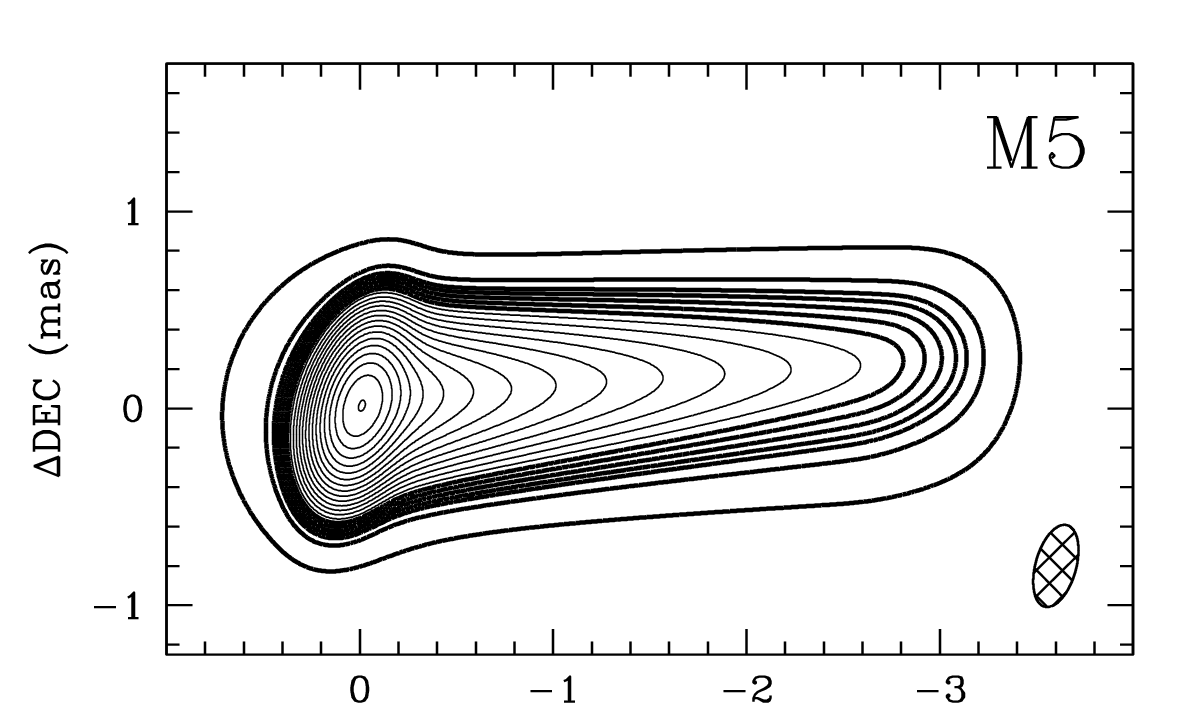}\\
\begin{minipage}[r]{0.48\textwidth}
  \includegraphics[width=\textwidth]{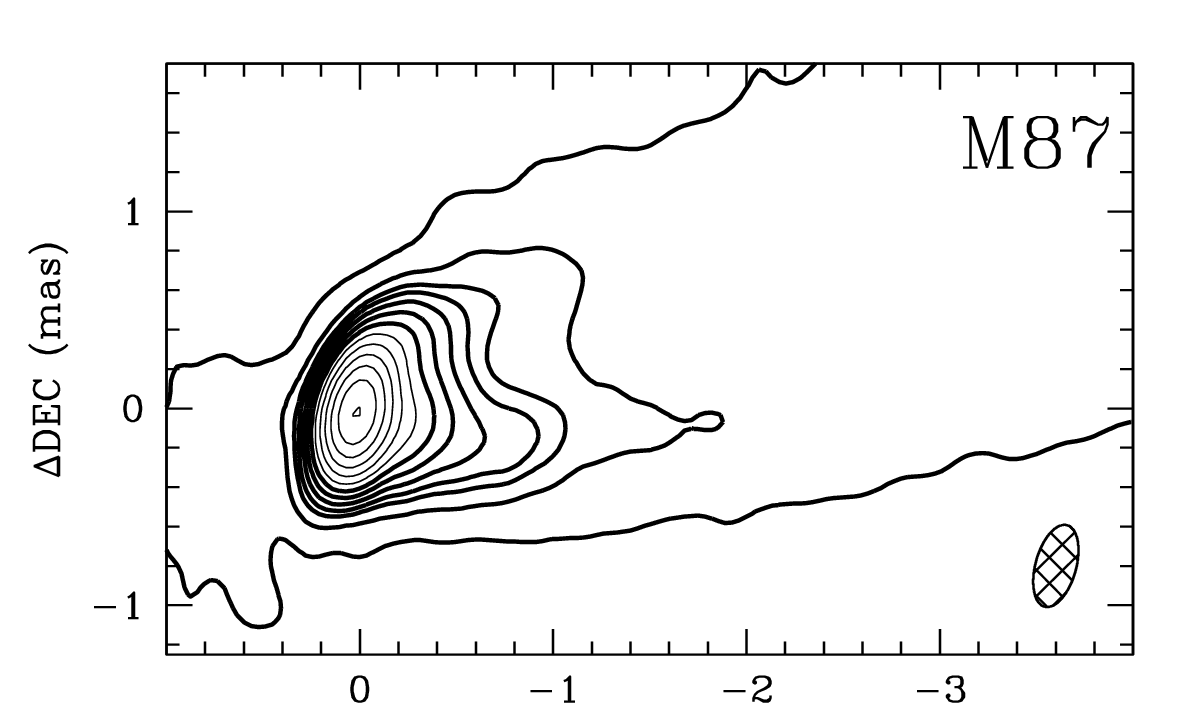}
\end{minipage}
\begin{minipage}[c]{0.48\textwidth}
  \caption{Above: $7\,\mm$ ($44\,\GHz$) images for the models listed in Table
    \ref{tab:models} and described in \S\ref{sec:MM.F}, convolved with
    the beam described in \citet{Walk-Ly-Juno-Hard:08} (shown in the
    lower-right).  The contours are at 0.1, 1, 2, 2.8, 4, 5.7 and 8
    mJy/beam (thick lines), and spaced in factors of $\sqrt{2}$ thereafter
    (thin lines).  In all figures, the
    jet is oriented along the x-axis (East-West), and the black hole
    is located at the origin.  The counter-jet is not visible in any
    of the images because it moves away from the observer and lacks
    the relativistic beaming of the forward jet.  These should be
    compared with the $7\,\mm$ image at left, produced from the
    average of 10 images taken througout 2007 and described in
    \citet{Walk-Ly-Juno-Hard:08} \citep[see also][]{Ly-Walk-Wrob:04}.
    For reference, $1\,\mas$ corresponds to roughly $220$
    Schwarzschild radii.\label{fig:7mm}}
\end{minipage}
\end{center}
\end{figure*}

In our model, the accreting plasma is characterized by both thermal
and non-thermal populations of electrons.  Both of these are
dynamically unimportant, with the underlying ions providing the
dominant pressure and determining the magnetic field strength.  The
particle densities are taken to be broken power laws, with a break at
some radius $R_{\rm b}$ which was subsequently adjusted to fit the
radio spectrum cut-off, making this qualitatively similar to the
truncated disk solutions discussed by \citet{Yuan:00}.  Outside of
$R_{\rm b}$, the power law indicies of the thermal particles were
taken from \citet{yuan03}, leaving only the normalizations to be
adjusted.  To fit the radio spectral index, we also modified the
radial dependence of non-thermal particles.

Specifically, the density of the thermal electrons is given by
\begin{equation}
n_{\rm th} = n_{\rm th,0} \,\e^{-z^2/2R^2} \left\{
\begin{array}{cc}
\displaystyle \left(r/R_{\rm b}\right)^{-0.7}
&
\text{if $r\ge R_{\rm b}$}\\
\displaystyle 1
&
\text{otherwise.}
\end{array}
\right.
\end{equation}
where $z$ is the height, $R$ is the cylindrical radius, and
$r=\sqrt{z^2+R^2}$.  The thermal electron temperatures is given by 
\begin{equation}
T_{e} = T_{e,0}\left(r/R_{\rm b}\right)^{-0.84}\,,
\end{equation}
which is different from the ion temperature, assumed to be at the
virial value.  The corresponding magnetic field strength is determined
via
\begin{equation}
\frac{B^2}{8\pi} = \beta^{-1} n_{\rm th} \frac{m_p c^2}{6 r}\,, 
\end{equation}
and oriented toroidally, where all lengths are measured in units of
$GM/c^2$.  In all models, $\beta=10$.

Similarly, the non-thermal electron density is given by
\begin{equation}
n_{\rm nth} = n_{\rm nth,0} \,\e^{-z^2/2R^2} \left\{
\begin{array}{cc}
\displaystyle \left(r/R_{\rm b}\right)^{-2} 
&
\text{if $r\ge R_{\rm b}$}\\
\displaystyle 1
&
\text{otherwise,}
\end{array}
\right.
\end{equation}
where the radial power-law index was chosen to roughly reproduce the
radio spectrum.  The associated spectral index is $3$ (i.e., the
spectral flux of the optically thin disk scales with frequency $\nu$
as $F_\nu \propto \nu^{-3}$).  The minimum Lorentz factor was set to
$100$, in approximate agreement with \citet{yuan03}.  Finally, the
disk velocity was assumed to be Keplerian outside of the ISCO and to
follow ballistic plunging inside.

As discussed in \S\ref{sec:MM.F}, qualitatively reproducing the 7mm
observations of Sgr A* requires the disk to become subdominant by this
wavelength.  We found that this requires $R_{\rm b} \simeq 20\,
GM/c^2$.  The normalizations for the thermal electrons are
$n_{\rm th,0} = 1.23\times10^4 \,\cm^{-3}$ and
$T_{e,0}=8.1\times10^9\,\K$.  The normalization for the non-thermal
electron density depends upon the particular model being considered
(see Table \ref{tab:models}).  For the case where the disk is viewed
from $40^\circ$ from the jet-axis, we set
$n_{\rm nth,0}=6.1\times10^{2}\,\rm cm^{-3}$, while for all other
cases $n_{\rm nth,0}=3.8\times10^{2}\,\rm cm^{-3}$.  Note that for all
models, the non-thermal electron density is substantially lower than
its thermal counterpart.

The associated disk spectrum is shown by the red, short-dashed line in
Fig \ref{fig:spec}.  As required, it becomes sub-dominant near
$7\,\mm$, and completely negligible by $1\,\mm$.  For this reason, the
particular details of the disk modeling are unlikely to affect the
millimeter wavelength images of M87's jet.

\subsection{Magnetically Dominated Jet} \label{sec:MM.J}
While a variety of mechanisms for the formation of ultra-relativistic
outflows have been proposed, the most successful are the magnetically
driven models.  Despite considerable uncertainty regarding their
detailed structure, magnetic jets share a number of general features,
which we seek to reproduce here.
These include providing a qualitatively realistic
relationship between the acceleration and collimation, the presence of
helical motion, the conversion from a poloidally dominated magnetic
field to a toroidal magnetic field at large distances, and a
transition from a jet to a trans-relativistic disk wind at large
values of the cylindrical radius.  Finally, any such model must be
numerically well behaved everywhere. 

In practice, all of these conditions could be met by appropriating
the output of recent jet-formation simulations
\citep[e.g.,][]{McKi-Blan:08,Tche-McKi-Nara:08,Igum:08,McKi:06,Hawl-Krol:06,Nish_etal:05}.
While this approach will undoubtedly prove productive in the future,
it is poorly suited to the goals of this investigation.  Presently,
despite being self-consistent, simulations are expensive, do not
provide sufficiently fine-grained controls over the jet parameters and
would not allow for comparisons outside the realm of MHD jets.  In
addition, there is considerable uncertainty regarding the
mass-loading, critical to the creation of images.  For these reasons,
we opted to produce a qualitatively correct cartoon of magnetically
dominated jets based upon force-free models of
\citet{Tche-McKi-Nara:08}, exhibiting many of the required properties
but providing sufficient freedom to explore a variety of jet
morphologies that may be outside of the purview of existing General
Relativistic MHD simulations.  Here we only summarize the physical
reasoning and the final structure of the jet, leaving many of the
details to the Appendix.

Numerical simulations suggest that outside of the accretion disk the
plasma is magnetically dominated \citep[see,
e.g.,][]{McKi-Blan:08,McKi:06,Hawl-Krol:06}.  Thus, we model the outflow as
stationary, axisymmetric and force-free.  In such configurations, the
magnetic field is described by a stream function, $\psi$, which must
satisfy a 2nd-order partial-differential equation.  A physically
relevant approximate solution is
\begin{equation}
\psi = r^{2-2\xi}\left(1 - \cos\theta\right)\,.
\end{equation}
For $0<\xi<1$, this solution is within $10\%$ of the stream function
inferred from existing numerical simulations
\citep{Tche-McKi-Nara:08}.  At large distances, $R \propto z^\xi$, and
thus $\xi$ defines the collimation rate.  For $\xi=0$ and $\xi=1$, the
jet is cylindrical and conical, respectively; a more typical value is
$1/2$.

The structure of the magnetic field and the velocity of the outflow
depend upon the angular velocity at the field-footprint,
$\Omega(\psi)$, which is a function solely of $\psi$.  The associated
four-velocity is
\begin{equation}
u^\mu_{F} = u^t_{F} \left( 1, 0, 0, \Omega(\psi) \right)\,,
\end{equation}
where $u^t_{F}$ is determined via a normalization condition.  Near
the light cylinder this choice of $u^\mu_{F}$ transitions from
time-like to space-like.  There is no physical problem with this
choice, meaning only that we have chosen a pathological observer.
Thus, we choose $u^t_{F}>0$ and set $\sigma=u^\mu_{F} u_{{\rm
F}\,\mu}=\pm 1$.  At the light cylinder, this is ill-defined.
However, as shown below, the plasma structure is continuous across
this surface, and thus we simply interpolate across the light cylinder
to obtain its characteristics at this position.  We define
$\Omega(\psi)$ in the equatorial plane, choose it to be the Keplerian
angular velocity outside of the ISCO, and fix it to the ISCO value
inside.  This boundary is usually what demarcates the ``jet'',
separating it from the surrounding disk wind.

\begin{figure*}
\begin{center}
\includegraphics[width=0.3\textwidth]{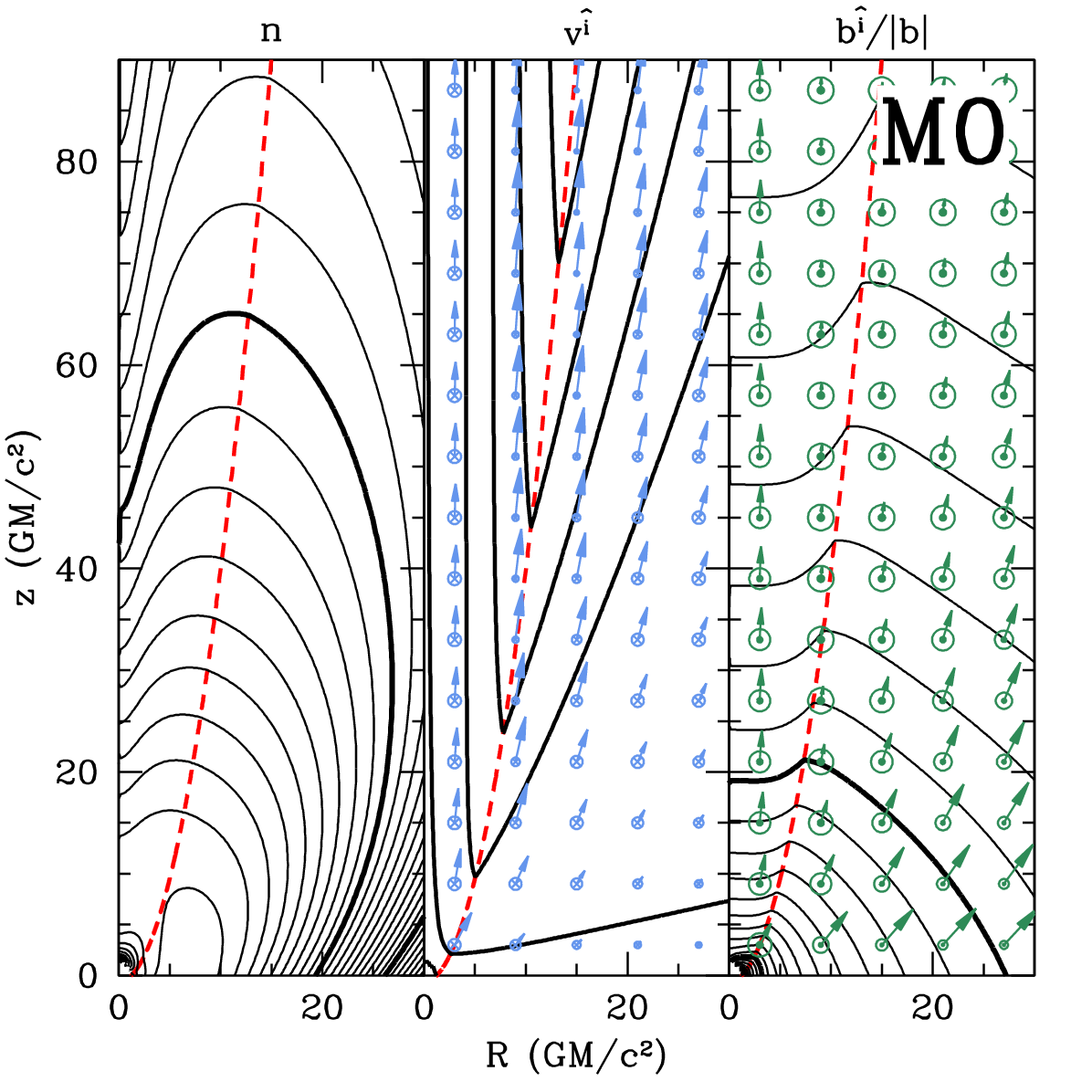}
\includegraphics[width=0.3\textwidth]{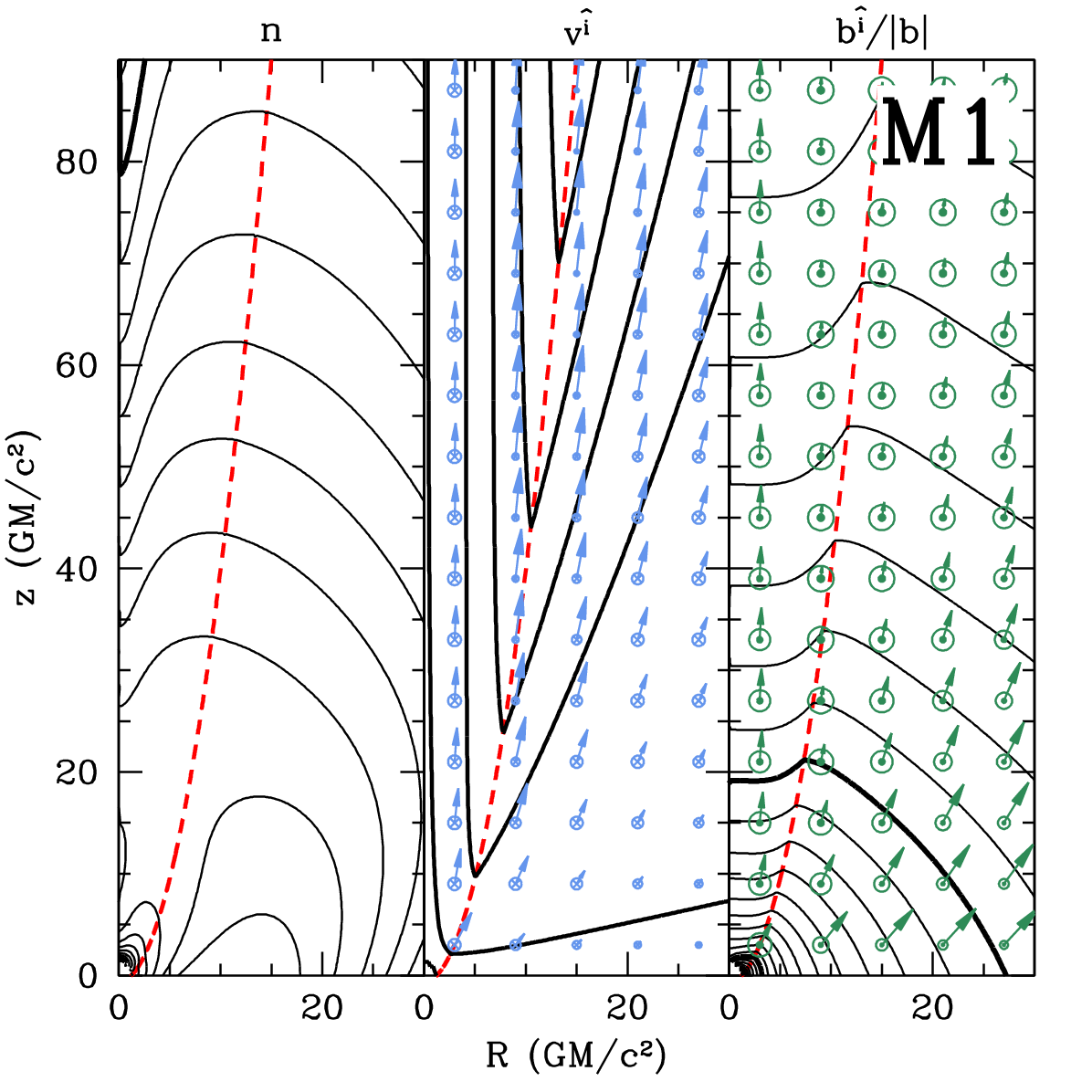}
\includegraphics[width=0.3\textwidth]{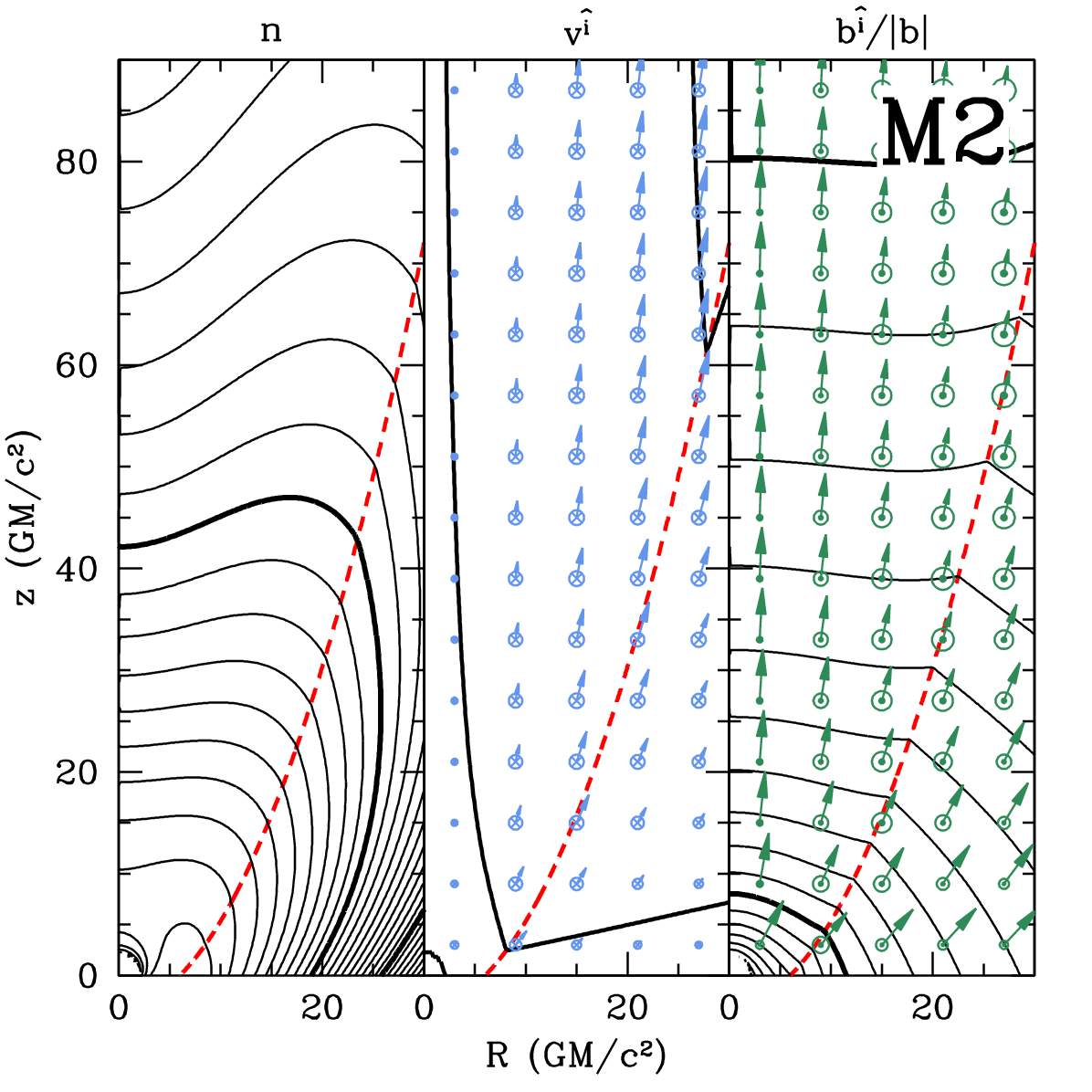}\\
\includegraphics[width=0.3\textwidth]{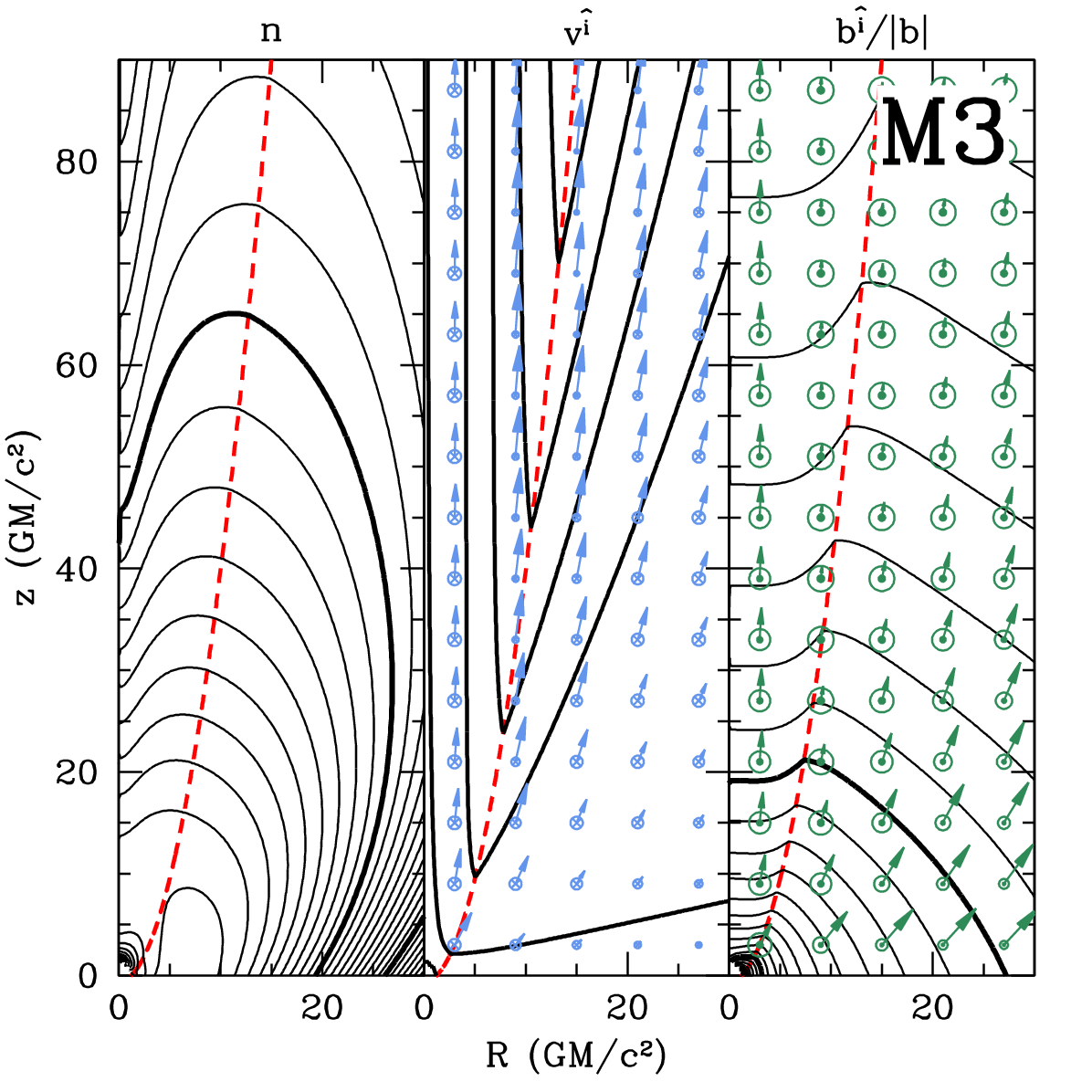}
\includegraphics[width=0.3\textwidth]{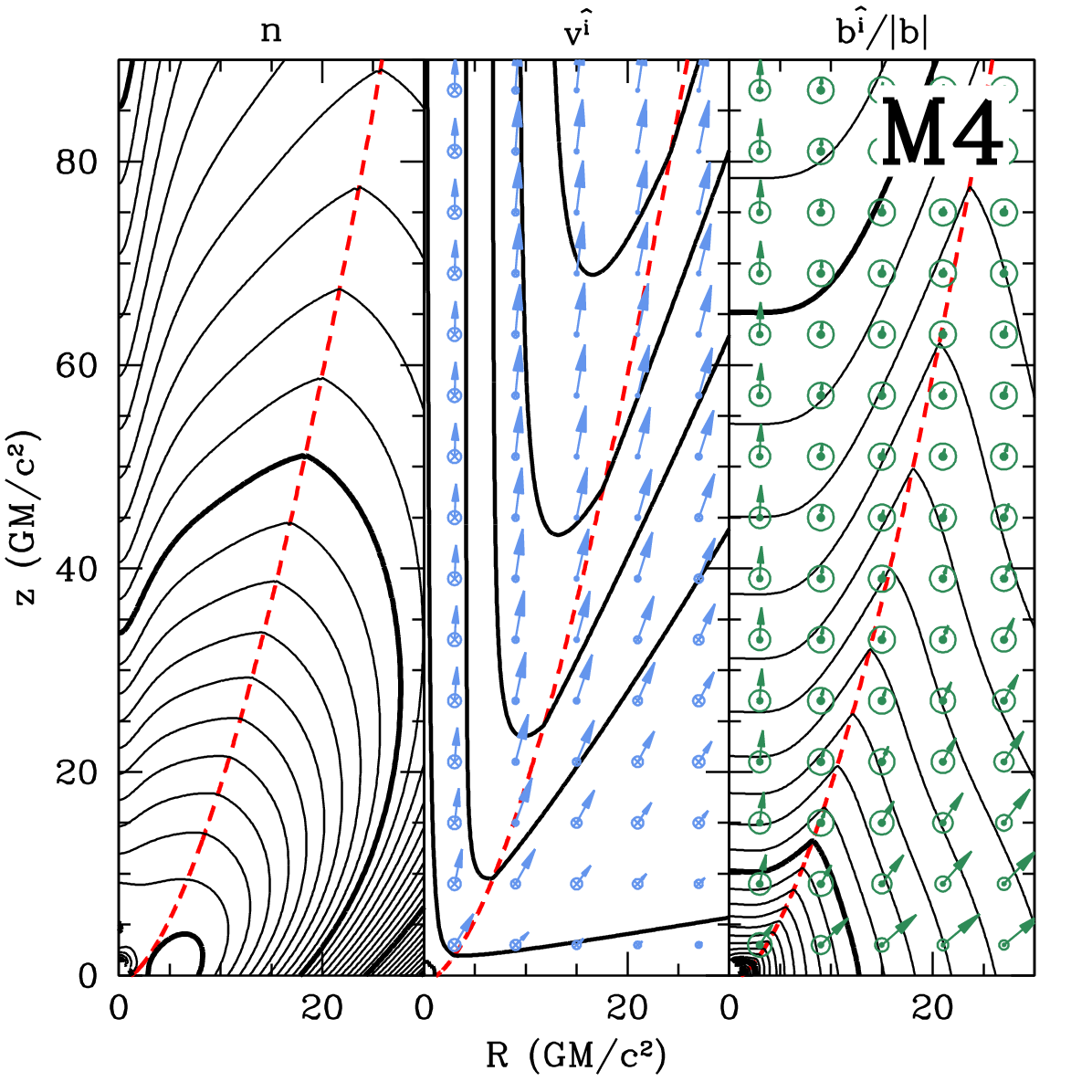}
\includegraphics[width=0.3\textwidth]{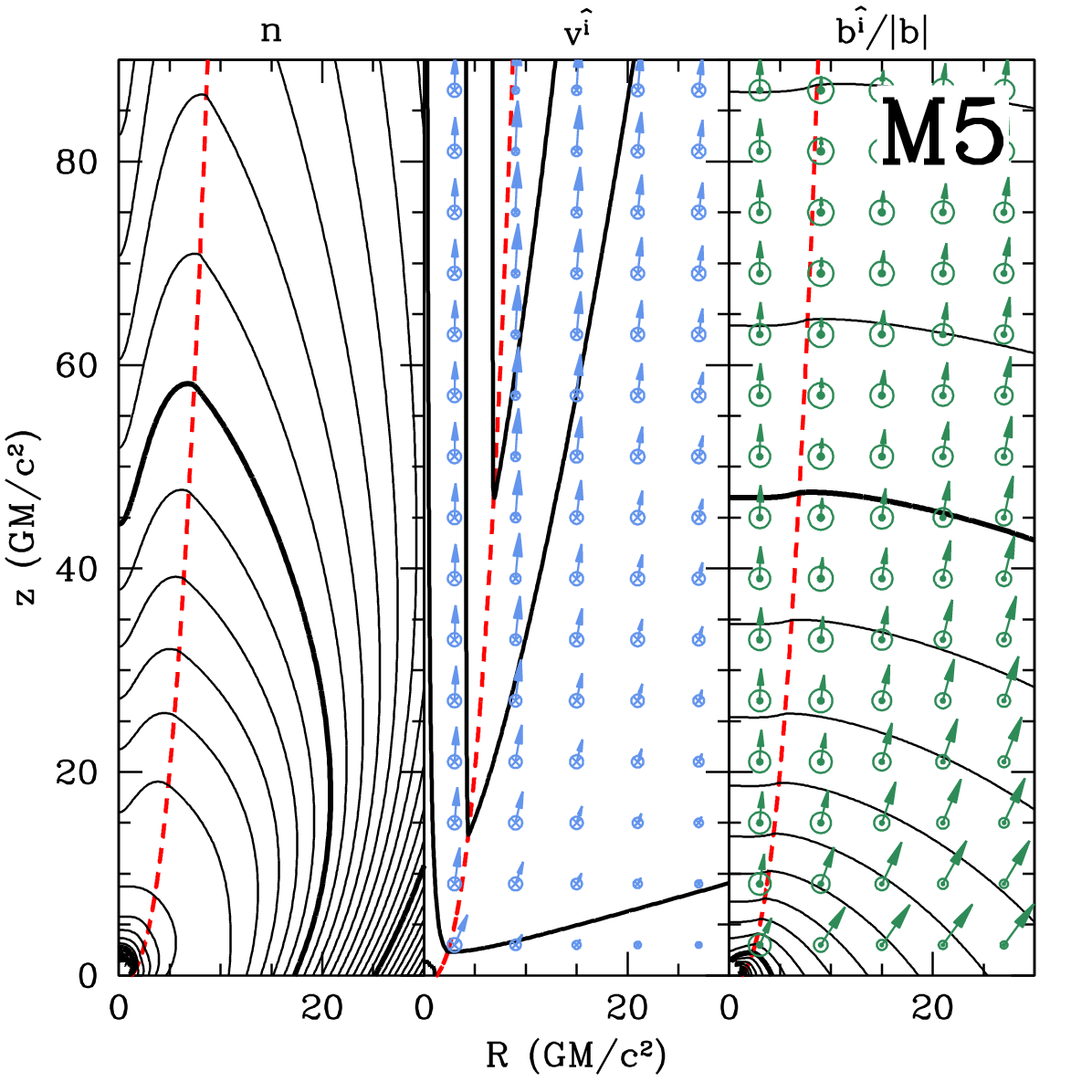}
\end{center}
\caption{Density, velocity and magnetic structure for the jet models
  listed in Table \ref{tab:models}.  For each, the density is shown in
  the left panel with thin/thick contours for
  $\log \left(n/n_{\rm max}\right)$ in steps of $0.1$/$1$.  The
  velocity is shown in the center panel, $v^{\hat{i}} = \sqrt{g_{ii}}
  u^i/u^t$, with contours of the asymptotic Lorentz factor ($-u_t$) in
  steps of $1$ (unit asymptotic Lorentz factor is shown by the lowest
  contour that extends across the entire range in $R$).  The magnetic
  field structure is shown in the right panel.  The contours denote
  the magnitude of the magnetic field in the plasma frame, and are
  distributed in the same fashion as the density contours.  Due to the
  considerable change in magnitude over the range shown, the magnetic
  field direction is shown by the vector field, where
  $b^{\hat{i}}=\sqrt{g_{ii}} b^i$.
  For both the velocity and magnetic field plots, the toroidal and
  poloidal vectors are normalized relative to each other such that they
  are equal when the toroidal circle has a diameter equal to the
  length of the poloidal vector.
  In all panels, the magnetic field surface passing through the ISCO
  is shown by the dashed red line.
} \label{fig:struc}
\end{figure*}

The stream function also defines the magnetic field structure.  Within
the $u_{F}$-frame (which we shall denote by the subscript $F$), the poloidal magnetic field 4-vector (see the Appendix for the
definition in terms of the electromagnetic field tensor) is given by
\begin{equation}
\begin{array}{l}
\displaystyle%
b_{F}^r
=
- B_0 \frac{\sigma}{u_{F}^t \sqrt{-g}} \frac{\partial \psi}{\partial \theta}
=
-B_0 \sigma \frac{r^{2-2\xi} \sin\theta}{u_{F}^t \sqrt{-g}}\\
\displaystyle%
b_{F}^\theta
=
B_0 \frac{\sigma}{u_{F}^t \sqrt{-g}} \frac{\partial \psi}{\partial r}
=
B_0 (2-2\xi) \sigma \frac{r^{1-2\xi} \left(1-\cos\theta\right)}{u_{F}^t \sqrt{-g}}
\end{array}
\end{equation}
The toroidal magnetic field may be constructed by noting that
$b_{{F}\, \phi} / u_{F}^t$ must be a function of $\psi$ alone
as well.  Choosing this function amounts to choosing the current
distribution within the outflow.  However, the condition that the
poloidal field dominates at large distances gives
\begin{equation}
b_{{F}\,\phi} = - 2 B_0 \Omega \psi u_{F}^t\,,
\end{equation}
where the minus sign arises from the fact that field lines are swept
back by the motion of the accretion disk.  Finally, $b_{F}^t$ is
obtained from the identity $u_{F}^\mu b_{{F}\,\mu} = 0$.

We assume that the plasma velocity is similar to the drift velocity as
seen in the frame of a zero-angular momentum observer (ZAMO).  This is
equivalent to assuming that the plasma is not relativistic at the jet
footprint within this frame; it gives
\begin{equation}
u^\mu = \gamma \left( u_{F}^\mu + \beta b_{F}^\mu \right)
\end{equation}
where
\begin{equation}
\gamma = -\frac{\sigma}{\sqrt{-(\sigma+\beta^2 b_{F}^2)}}
\quad\text{and}\quad
\beta = \frac{\sigma b_{F}^t}{b_{F}^2 u_{F}^t}\,.
\end{equation}
As shown in the Appendix, unlike $u_{F}^\mu$ this velocity is
always time-like.  The magnetic field 4-vector in the plasma frame is
then
\begin{equation}
b^\mu = \gamma \left( b_{F}^\mu - \sigma\beta b_{F}^2 u_{F}^\mu \right)\,.
\end{equation}

All that remains is to define the mass content of the outflow.  To do
this properly would require a full understanding of the mass-loading
and particle acceleration in jets, that presently does not exist.
Therefore, we employ a simple, parametrized prescription based upon
mass conservation.  As shown in the Appendix, the particle density $n$
satisfies in Boyer-Lindquist coordinates,
\begin{equation}
\frac{\gamma n}{u_{F}^t b_{F}^2} \left(g^{tt} +
g^{t\phi}/\Omega\right) = F(\psi)\,,
\end{equation}
for an arbitrary function of the stream function.  We choose the
particular form of $F(\psi)$ by defining the jet density $n$ on a
slice at some height above the disk.  Specifically, we chose the
profile to be a Gaussian of the form $n\propto
\exp\left(-r^2\sin^2\theta/2 r_{\rm fp}^2\right)$, at a height
$|r\cos\theta| = r_{\rm fp}$.  With $b_{F}^\mu$, $u_{F}^t$ and
$\Omega$ known, this is sufficient to define $F(\psi)$
everywhere. Finally, this was supplemented with a cut-off for $r <
r_{\rm fp}$, mimicking the loading of the jet above the black hole,
yielding
\begin{equation}
n = n_0 \frac{u_{F}^t b_{\rm
F}^2F(\psi)}{\gamma\left(g^{tt}+g^{t\phi}/\Omega\right)} \left( 1 -
\e^{-r^2/2 r_{\rm fp}^2}\right)\,.
\end{equation}
The shape of the non-thermal electron distribution within jet cores is
presently the subject of considerable debate.  Even within the context
of power-law models, which we employ here, observational estimates of
the minimum Lorentz factor can vary from $10$ to $10^5$ depending upon
the object, emission model and type of observation
\citep[see, e.g.,][]{Falc-Bier:95,Pian_etal:98,Kraw-Copp-Ahar:02,Jors-Mars:04,Soko-Mars:05,Saug-Henr:04,Kata_etal:06,Gieb-Dubu-Khel:07,Muel-Schw:08,Stee-Blun-Duff:08}.
For concreteness, similar to the disk we choose the minimum Lorentz
factor to be $100$, which is broadly consistent with the spectral
characteristics of M87, as we will see in the following section.  In
contrast to the disk, however, we choose a spectral index between
$1.0$ and $1.13$, depending upon the particular model in Table 1.

In summary, the parameters that control the jet are the collimation
rate, $\xi$, the footprint size, $r_{\rm fp}$, the magnetic field and
density normalization, $B_0$ and $n_0$.  The latter two are set by
fitting M87's spectrum, leaving the first two unconstrained.

\subsection{Fitting Existing Constraints} \label{sec:MM.F}
The two empirical constraints that we considered were M87's SED and the
$7\,\mm$ VLBI images.  These are not completely independent since, as
alluded to earlier, the jet at $7\,\mm$ must be sufficiently
bright to produce the extended emission that is observed in the VLBI
images.  Conversely, this requires that the disk become sub-dominant,
and thus is the motivation behind the truncated disk.  In addition,
the cut-off in the spectrum due to the minimum Lorentz factor of the
non-thermal jet electrons is critical to forcing the jet component of
the spectrum to turn over near $1\,\mm$, ensuring that it does not
over-produce the flux at longer wavelengths.

The particular normalizations for the electron densities and magnetic
fields in the disk and jet are determined by fitting the SED shown in
Fig. \ref{fig:spec}.  The disk component is shown by the red,
short-dashed line and falls off rapidly near $7\,\mm$.  The jet
component for our conical jet model (M0, Table \ref{tab:models}) is
shown by the blue, long-dashed line.  This fits the
millimeter--optical spectrum, and is dominated by the disk spectrum
above $7\,\mm$.  Within this model there is some detailed balancing of
the two components, which is presumably associated with the physics of
jet formation.  

Despite treating the disk and jet components independently, making no
attempt to merge the two, we find that the jet magnetic field and the
magnetic field in the inner part of the disk are quite similar,
differing by factors of a few.  In contrast, the jet density is
typically considerably lower than disk densities, being an order of
magnitude smaller than that of the non-thermal electrons in the disk.
However, this disparity is consistent with jet simulations which have
generally shown that it is quite difficult for disk material to
diffuse onto the ordered field lines within the jet
\citep{McKi-Gamm:04,Komi:05,DeVi-Hawl-Krol-Hiro:05,McKi:06}.

\begin{deluxetable}{clccc}
\tablecaption{Jet model parameters\label{tab:models}}
\tablehead{
\colhead{Model} &
\colhead{$a (M)$} &
\colhead{$\theta$} &
\colhead{$r_{\rm fp} (GM/c^2)$} &
\colhead{$\xi$}
}
\startdata
M0 & 0.998 & $25^\circ$ & 10 & $1/2$\\
M1 & 0.998 & $25^\circ$ & 20 & $1/2$\\
M2 & 0 & $25^\circ$ & 10 & $1/2$\\
M3 & 0.998 & $40^\circ$ & 10 & $1/2$\\
M4 & 0.998 & $25^\circ$ & 10 & $5/8$\\
M5 & 0.998 & $25^\circ$ & 10 & $3/8$
\enddata
\end{deluxetable}

We consider six jet-disk models altogether, the parameters of which
are listed in Table \ref{tab:models}.  These explore the dependence of
the jet structure and images upon the variation of a single jet
parameter at a time.  The spectra for all models are nearly
indistinguishable from that shown in Fig. \ref{fig:spec}.  The
$7\,\mm$ image, convolved with the beam reported in
\citet{Walk-Ly-Juno-Hard:08} and \citet{Ly-Walk-Wrob:04}
\citep[see also][]{Juno-Bire-Livi:99} is shown for each model in
Fig. \ref{fig:7mm}.  Generally we find an acceptable qualitative
agreement, which is not trivial given the dubious extension of the
force-free jet model to these angular scales (corresponding to many
hundreds of Schwarzchild radii).  We note, however, that
the brightness profile of our canonical jet model decreases much more
rapidly ($\propto z^{-3}$) than that observed by
\citet{Walk-Ly-Juno-Hard:08} ($\propto z^{-1.8}$), implying that
corrections to our simplistic model are required far from the black
hole.  Indeed, because we don't model the instabilities and
subsequent internal shocks noted in simulations, which ultimately
redistribute the jet's magnetic and particle energy, our jet
continues to accelerate, reaching unphysically large Lorentz factors
by these angular scales.  Nevertheless, we obtain roughly the correct
morphology for the single jet orientation we consider, East-West.

The density, velocity and magnetic field structure, shown in
Fig. \ref{fig:struc} for the scale that dominates the millimeter
images, share a number of common features.  In all, the critical
surface defined by that poloidal field line which passes through the
ISCO, shown by the red dashed line, is associated with a qualitative
change in the outflow.  Following \citet{McKi:06}, by ``jet'' we will
refer the region inside this surface, and by ``wind'' we mean the
region outside.  It is along this surface that the highest Lorentz
factors are reached and the velocity is most poloidal (i.e., the
velocity pitch angle is largest).  As we move away from this surface,
both within the jet and the wind, the toroidal velocity component
increases.  Conversely, the magnetic field is most toroidal (i.e., the
magnetic pitch angle is smallest) along this surface, with an
increasing poloidal component on both sides.  Finally, the density
generically reaches a maximum along this surface.

This behavior is not unexpected, deriving directly from the velocity
of the magnetic field footprints.  Outside of the ISCO, the footprint
velocity is monotonically decreasing with radius (since $\Omega$ is
set to the Keplerian value).  Inside of the ISCO $\Omega$ is held
fixed and thus the velocity is increasing with radius.  Thus, the
critical magnetic field surface corresponds to the maximum footprint
velocity, with the jet and wind structure on either side roughly
mirroring each other.

\begin{figure*}
\begin{center}
\includegraphics[width=0.3\textwidth]{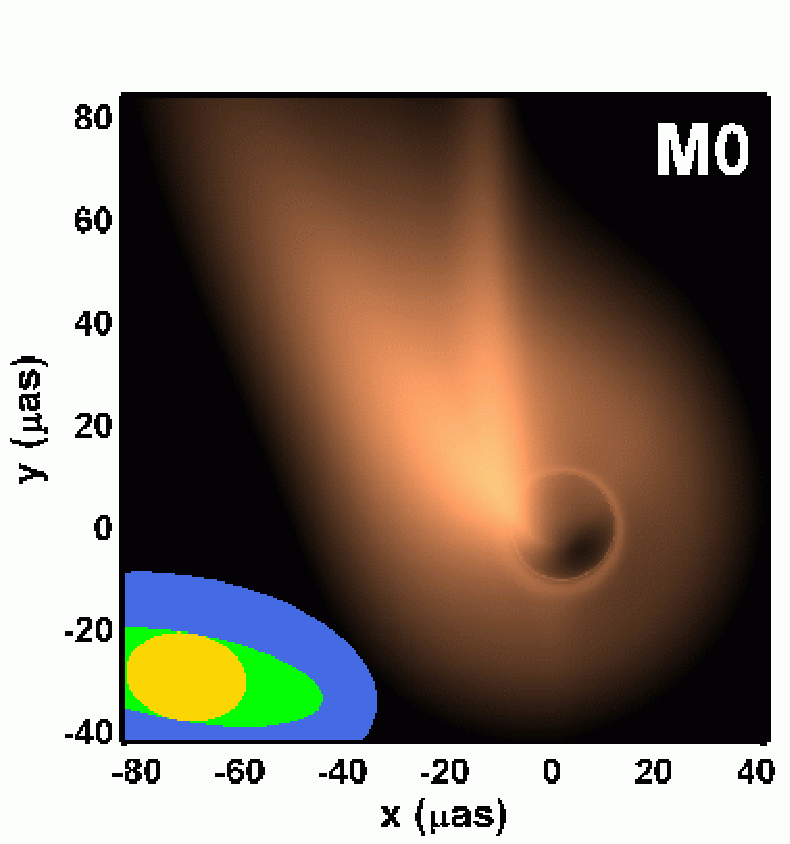}
\includegraphics[width=0.3\textwidth]{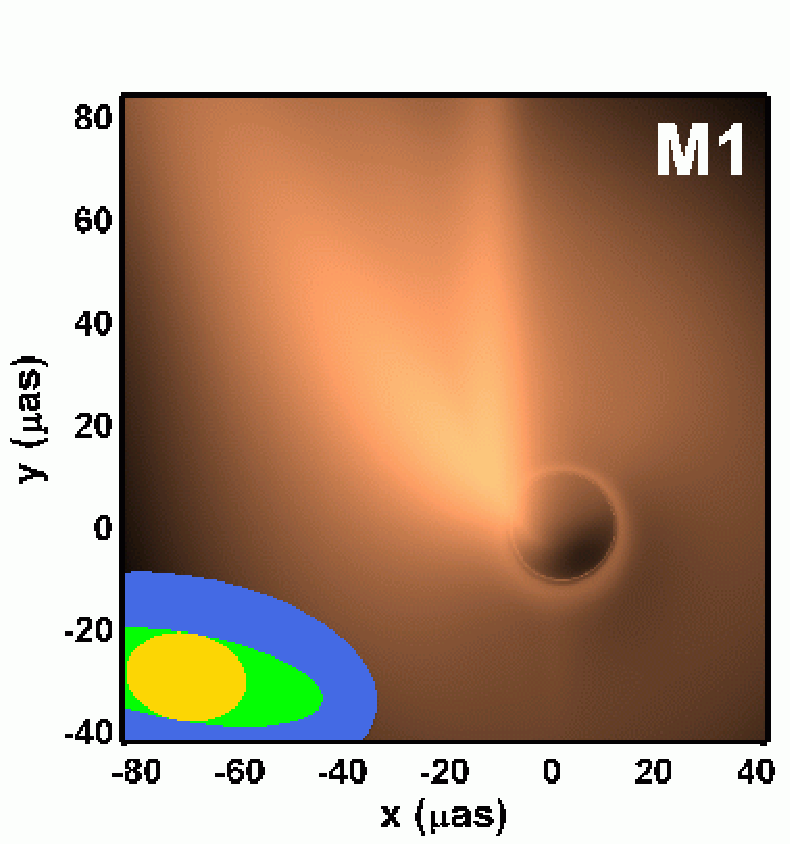}
\includegraphics[width=0.3\textwidth]{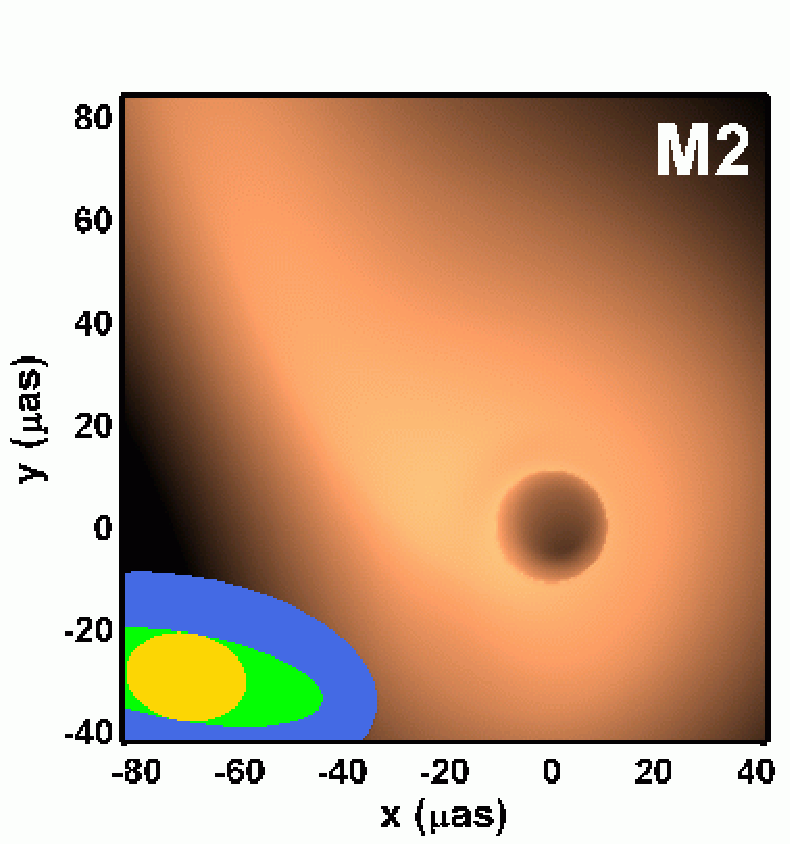}\\
\includegraphics[width=0.3\textwidth]{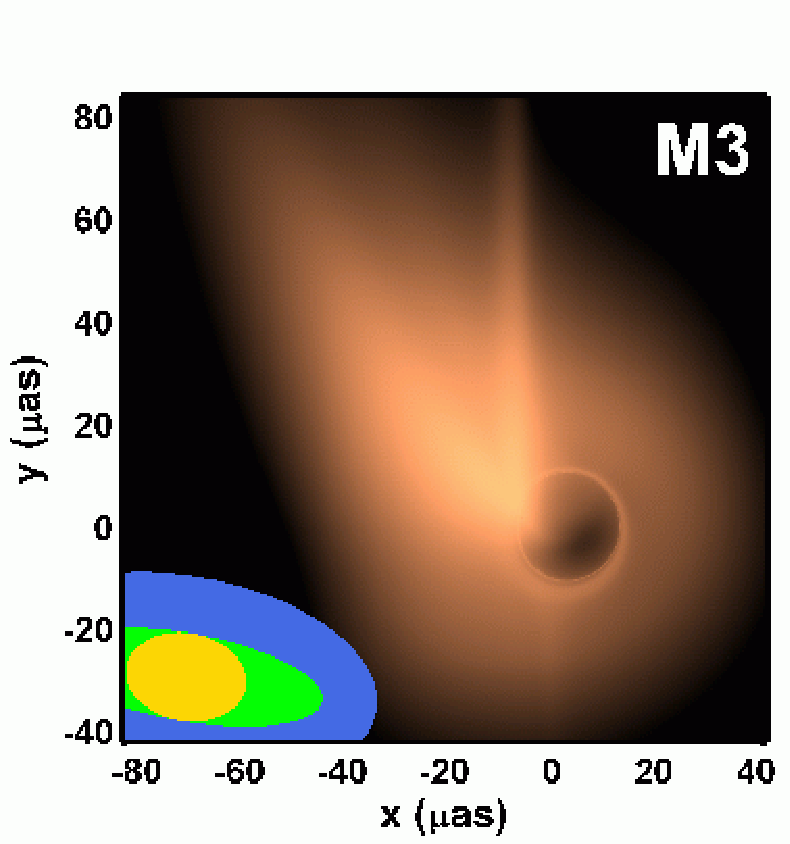}
\includegraphics[width=0.3\textwidth]{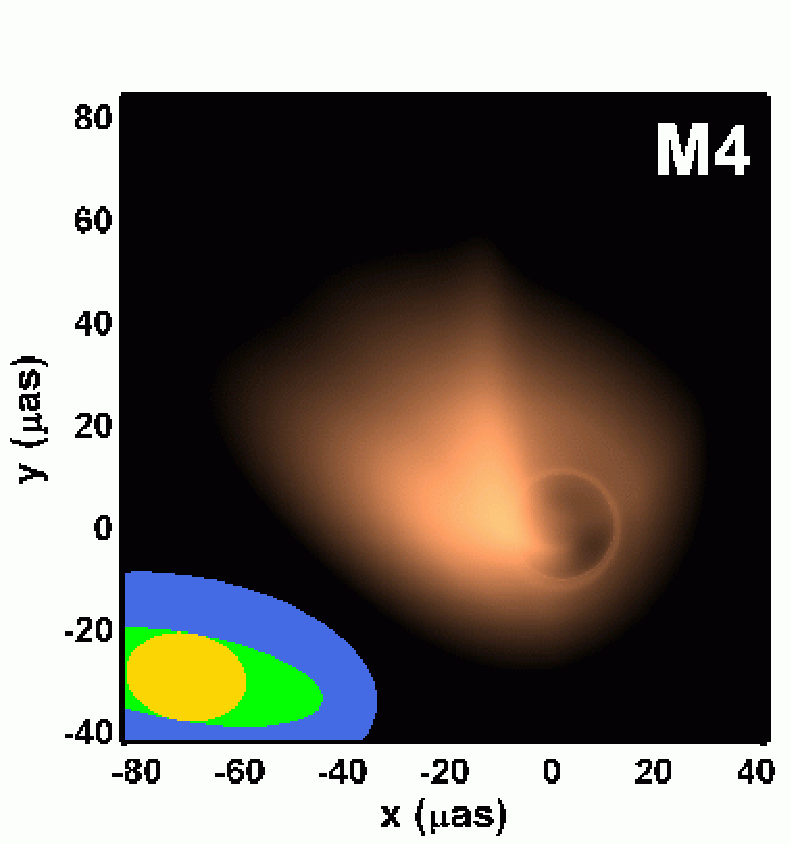}
\includegraphics[width=0.3\textwidth]{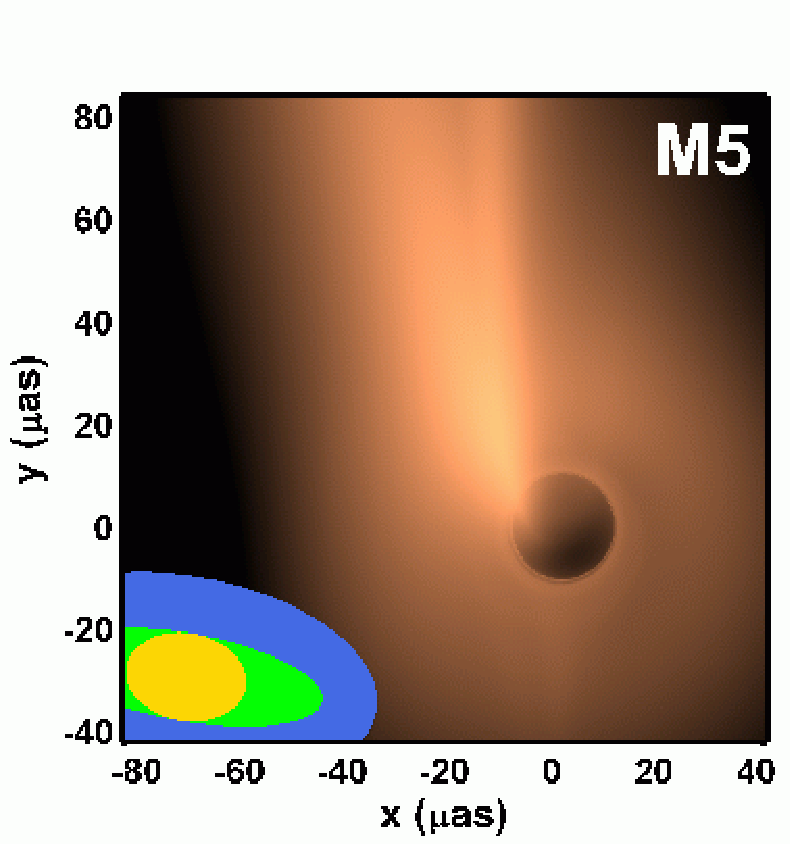}
\end{center}
\caption{$1.3\,\mm$ ($230\,\GHz$) images for the models listed in Table
  \ref{tab:models} and described in \S\ref{sec:MM.F}.  The brightness
  scale is logarithmic, showing a dynamic range of 128, similar to the
  existing $7\,\mm$ images.  In all images the projected jet axis is
  parallel to the vertical axis.  The ellipses in the lower left
  provide an estimate of the beam size for arrays consisting of the
  North American telescopes only (blue), including the European
  telescopes \& the LMT (green), and all telescopes (yellow).}
  \label{fig:230}
\end{figure*}

Next we discuss the individual model features separately.
\subsubsection{M0}
M0 represents our canonical jet model, having parameters that are are
broadly consistent with numerical simulations and direct observations
of M87.  In this case the black hole is rapidly rotating, has a
footprint of $r_{\rm fp}=10 GM/c^2$ 
and collimation index of $\xi=1/2$.  Spectra and images are produced
by viewing the jet at $25^\circ$ from the jet/spin axis, as inferred
from observations of super-luminal knots \citep{Hein-Bege:97}.  The
asymptotic Lorentz factor reaches 5 within the region shown in
Fig. \ref{fig:struc}, and is comparable to that inferred in M87
\citep{Hein-Bege:97}.  This model easily produces the extended
$7\,\mm$ emission, with the slight deviation from purely East-West
orientation arising due to the helical motion of plasma in the jet.

\subsubsection{M1}
M1 is identical to M0, save for the increased footprint size, $r_{\rm
fp}=20$.  This may be the case, e.g., if black hole spin does not play
the central (anticipated) role in launching the jet.  The velocity and
magnetic field structures are identical to those of model M0, the only
difference being the width of the density distribution.  This has
consequences for the normalizations of the density and magnetic field,
with the larger geometric area forcing these to be substantially
smaller in magnitude.  Again, this model is qualitatively able to
reproduce the $7\,\mm$ observations, perhaps even better than model
M0.

\subsubsection{M2}
M2 corresponds to a jet from a non-rotating black hole.  The primary
consequence of setting $a=0$ is to increase the ISCO radius, thus
substantially decreasing the maximum footprint velocity and widening
the critical magnetic surface.  Thus, even with $r_{\rm fp}=10$ and
$\xi=1/2$, as in M0, the jet appears considerably broader, and is
accelerated much more slowly.  In the region shown by
Fig. \ref{fig:struc}, the maximum asymptotic Lorentz factor reached is
only 2.  Nevertheless, M2 is capable of producing the extended
$7\,\mm$ emission that is observed.

\subsubsection{M3}
The structure of the jet in M3 is identical to M0, the only difference
being the viewing angle.  For M3, this is chosen to be $40^\circ$ from
the jet/spin axis.  As a consequence, the density and magnetic
normalizations of the jet and the underlying disk needed to be
adjusted.  Otherwise, there is no difference in the modelling.

\subsubsection{M4}
M4 is a moderately less collimated jet, with $\xi=3/8$.  This
relatively small change in the collimation rate has dramatic
consequences for the density and magnetic field structure, both
falling off much more rapidly in this case.  The shape of the
asymptotic Lorentz factor
contours are somewhat different from those associated with M0,
exhibiting a more gradual transition between the jet and wind
regions.  Nevertheless, the maximum asymptotic Lorentz factor reached
is 5, similar to M0.  In this case, the $7\,\mm$ image is considerably
different than that of the previous models, appearing more compact due
primarily to the rapid density fall-off with radius.

\subsubsection{M5}
Finally, M5 corresponds to a moderately more collimated jet, with
$\xi=5/8$.  Again this results in a substantial change in the density
and magnetic field structure, this time narrowing the jet.  Unlike M4,
the asymptotic Lorentz factor is also significantly altered, reaching
only 3 in the region shown.  Thus, the more rapid collimation results
in a more slowly accelerating, narrow jet.  This overproduces the
extent of the $7\,\mm$ emission, and has difficulty producing the
observed breadth.

\section{Jet Images} \label{sec:JI}

\begin{figure*}
\begin{center}
\includegraphics[width=0.3\textwidth]{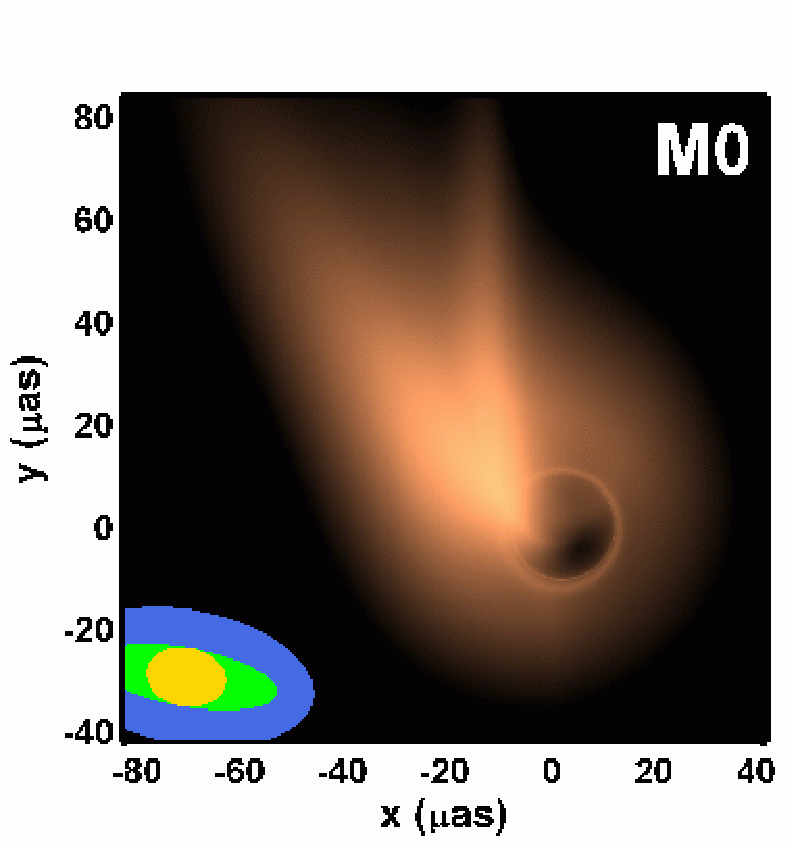}
\includegraphics[width=0.3\textwidth]{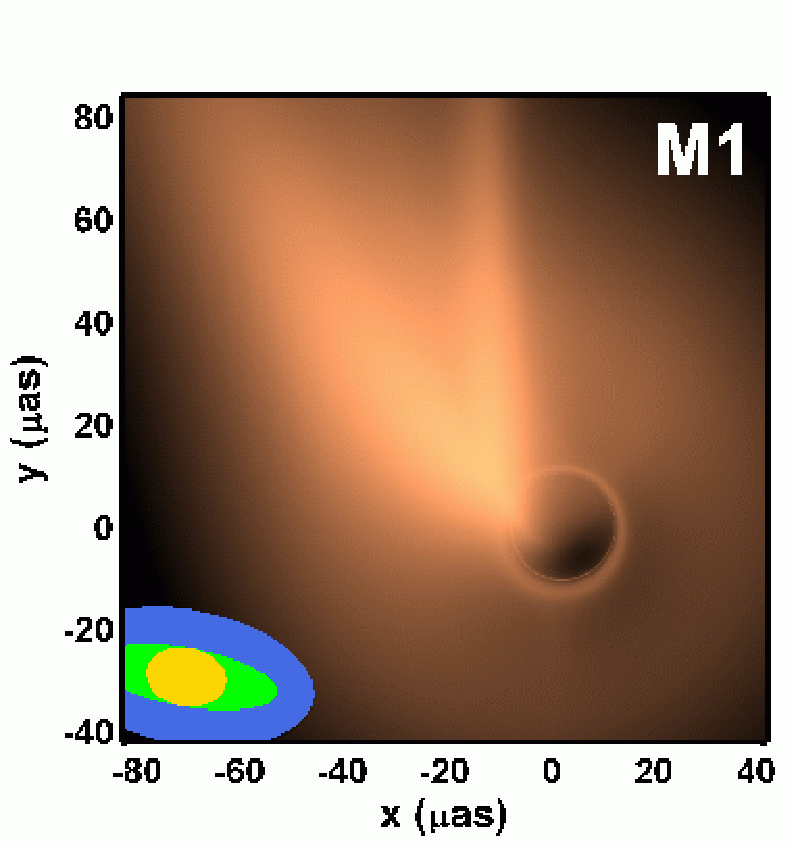}
\includegraphics[width=0.3\textwidth]{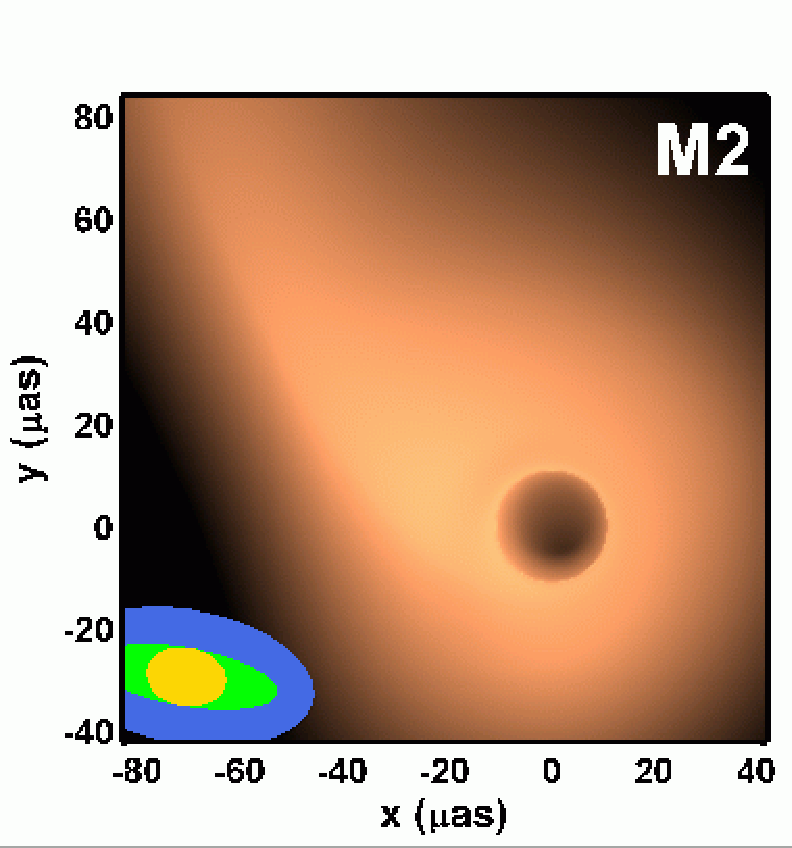}\\
\includegraphics[width=0.3\textwidth]{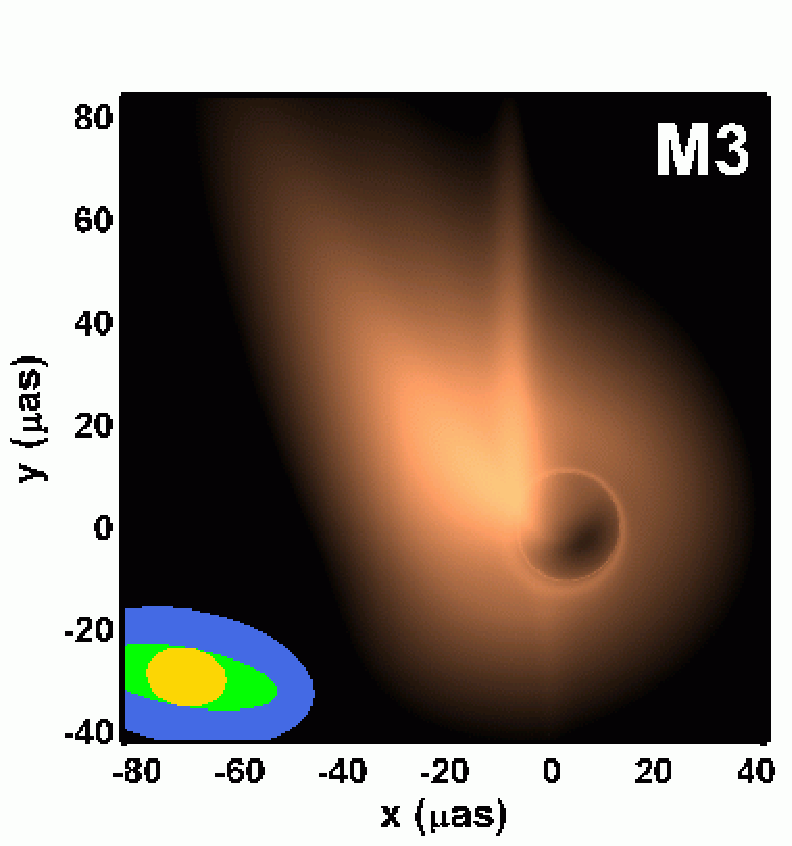}
\includegraphics[width=0.3\textwidth]{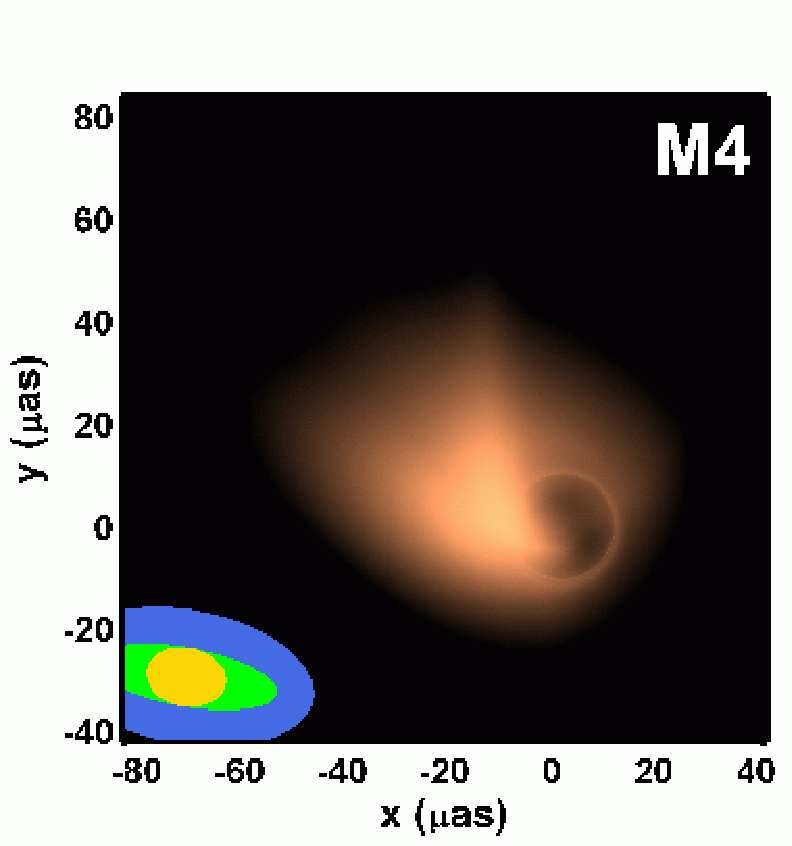}
\includegraphics[width=0.3\textwidth]{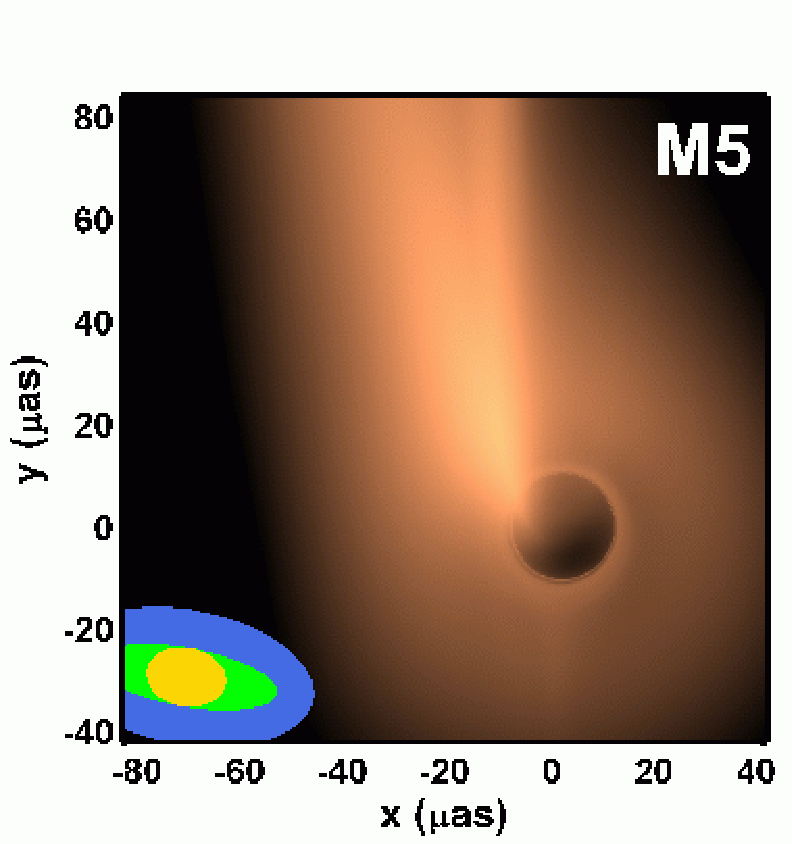}
\end{center}
\caption{$0.87\,\mm$ ($345\,\GHz$) images for the models listed in Table
  \ref{tab:models} and described in \S\ref{sec:MM.F}.  The brightness
  scale is logarithmic, showing a dynamic range of 128, similar to the
  existing $7\,\mm$ images.  In all images the projected jet axis is
  parallel to the vertical axis.  The ellipses in the lower left
  provide an estimate of the beam size for arrays consisting of the
  North American telescopes only (blue), including the European
  telescopes \& the LMT (green), and all telescopes (yellow).}
\label{fig:345}
\end{figure*}

\subsection{$1.3\,\mm$}

Images of M87 at $1.3\,\mm$ ($230\,\GHz$) are presented in
Fig. \ref{fig:230}.  As
with images of accretion flows \citep{Brod-Loeb:06a,Brod_etal:08}, the
outflow images are strongly influenced by relativistic beaming,
boosting and strong gravitational lensing.  In all images the jet is
lopsided, with the emission dominated by the side in which the
helically moving outflow is approaching the observer.  However, whereas
for accretion flows this results in a bright crescent, for our outflow
model this produces a bifurcated image.  This is clearly seen in the
$1.3\,\mm$ image of M0, showing bright diagonal and vertical branches,
which can be understood in terms of the generic velocity structure of
the jet.  As discussed in \S\ref{sec:MM.F}, the velocity is both
largest and nearly radial along the critical magnetic field surface
that divides the jet from the wind.  Here, relativistic beaming dims
the emission outside of a small angle, producing a flux deficit.  Both
inside and outside this surface, the toroidal component of the
velocity increases, and the velocity is aligned along the line of
sight, producing bright regions.  The bright vertical and diagonal
branches are due to emission from the jet and wind, respectively.

Even at a wavelength of $1.3\,\mm$, the jet is typically not completely
optically thin.  This is not unexpected given that the jet spectrum is
just turning over at $1.3\,\mm$.  This is most pronounced near the jet
base, which obscures a portion of the black hole horizon.
Nevertheless, in all images the silhouette cast by the black hole upon
the lensed image of the base of the counter-jet\footnote{In this case,
  we are seeing the counter-jet near to the black hole, where it has
  not yet accelerated appreciably and thus relativistic beaming is not
  yet effective.  Far from the black hole the jet is rapidly moving
  away from the observer and beaming renders it invisible, as seen in Fig
  \ref{fig:7mm}.  Should the jet accelerate much more rapidly than
  modeled here, or the counter-jet not be present, the silhouette will
  be much weaker.}
is clearly visible.
This suggests that imaging a black hole silhouette may be more easily
done in M87 than in Sgr A*.  However, at $1.3\,\mm$, the optimal beam,
obtained by phasing together an array including all existing North
American and European telescopes, the LMT and ALMA, is comparable in
extent to the silhouette.

For the purpose of informing jet-formation modelling, it is fortuitous
that each of the models is clearly distinguishable.  The wider
footprint of M1 results in a correspondingly broader brightness
distribution.  The lower velocities associated with lack of spin (M2)
produces a featureless, asymmetric image with a single branch (though
completely optically thin).  Changing the inclination (M3) produces a
broader base and narrower jet branch, though this case is most
degenerate with the image from the canonical model (M0).  It is not
obvious whether images of the quiescent jet emission alone will
distinguish these two cases.  On the other hand, altering the
collimation index even a small amount produces drastically different
millimeter images (M4 \& M5).  Thus, short-wavelength imaging promises
to severely constrain the jet footprint size, the black hole spin and
the collimation index, all providing critical observational input into
existing efforts to model the structure of MHD jets.

\begin{figure*}
\begin{center}
\includegraphics[width=0.3\textwidth]{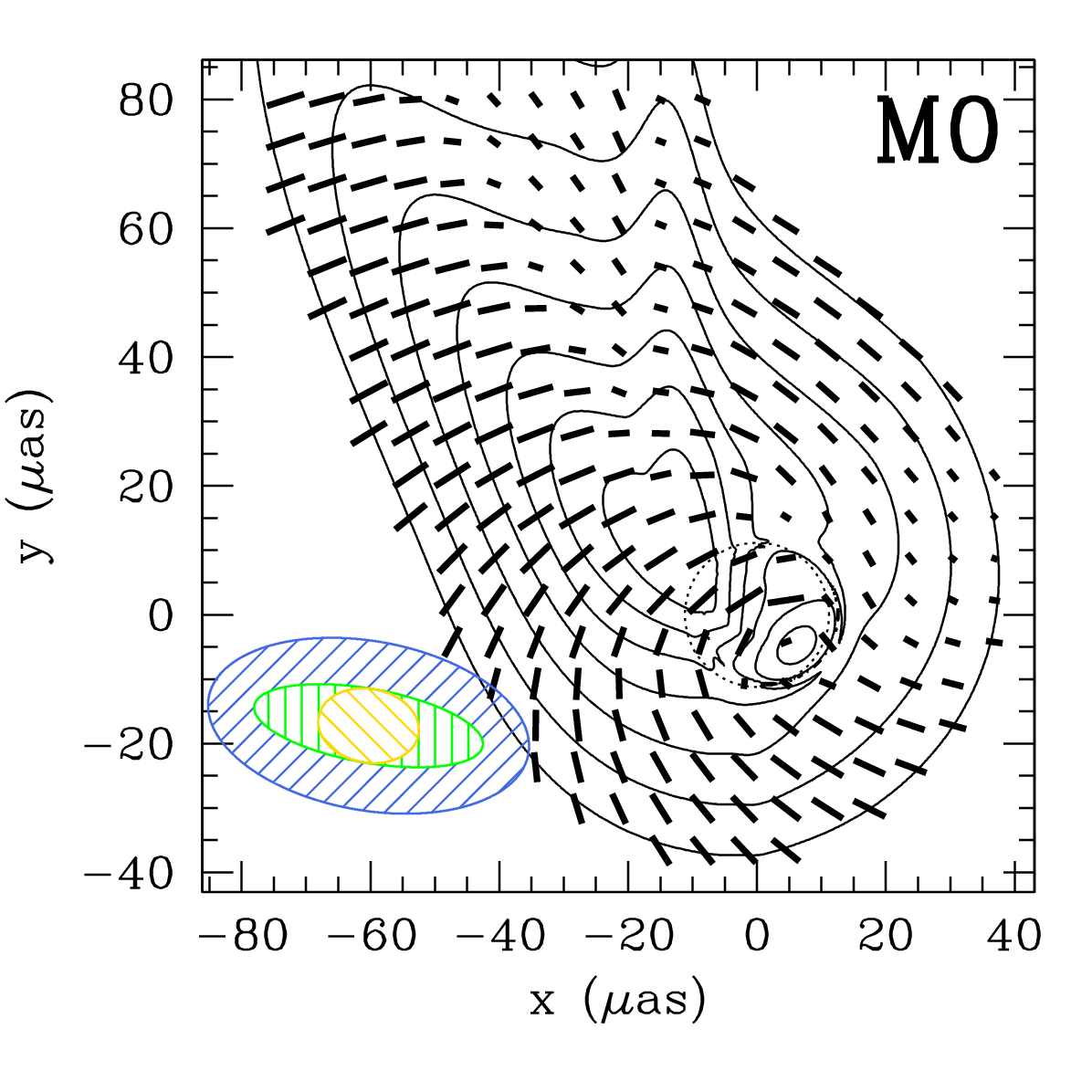}
\includegraphics[width=0.3\textwidth]{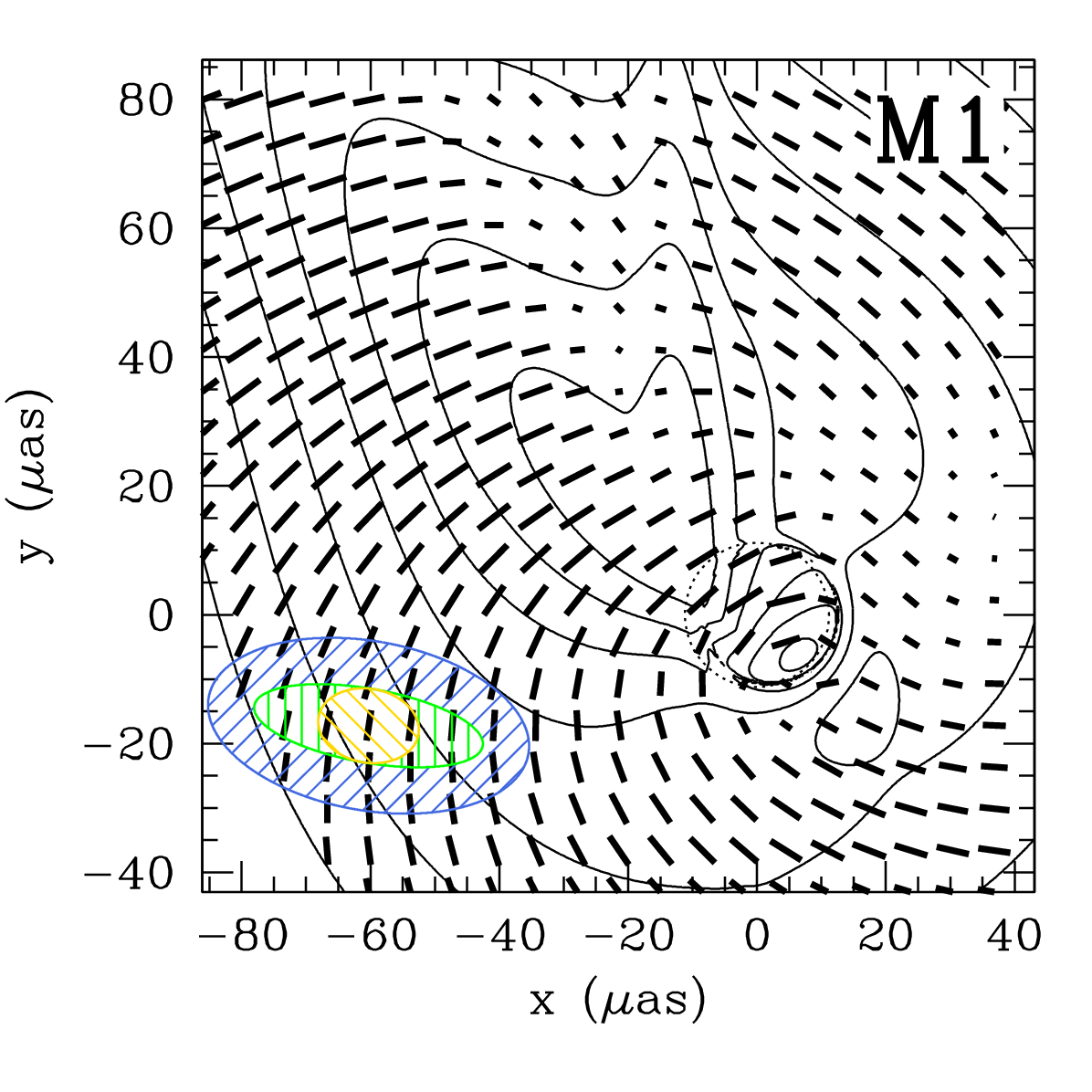}
\includegraphics[width=0.3\textwidth]{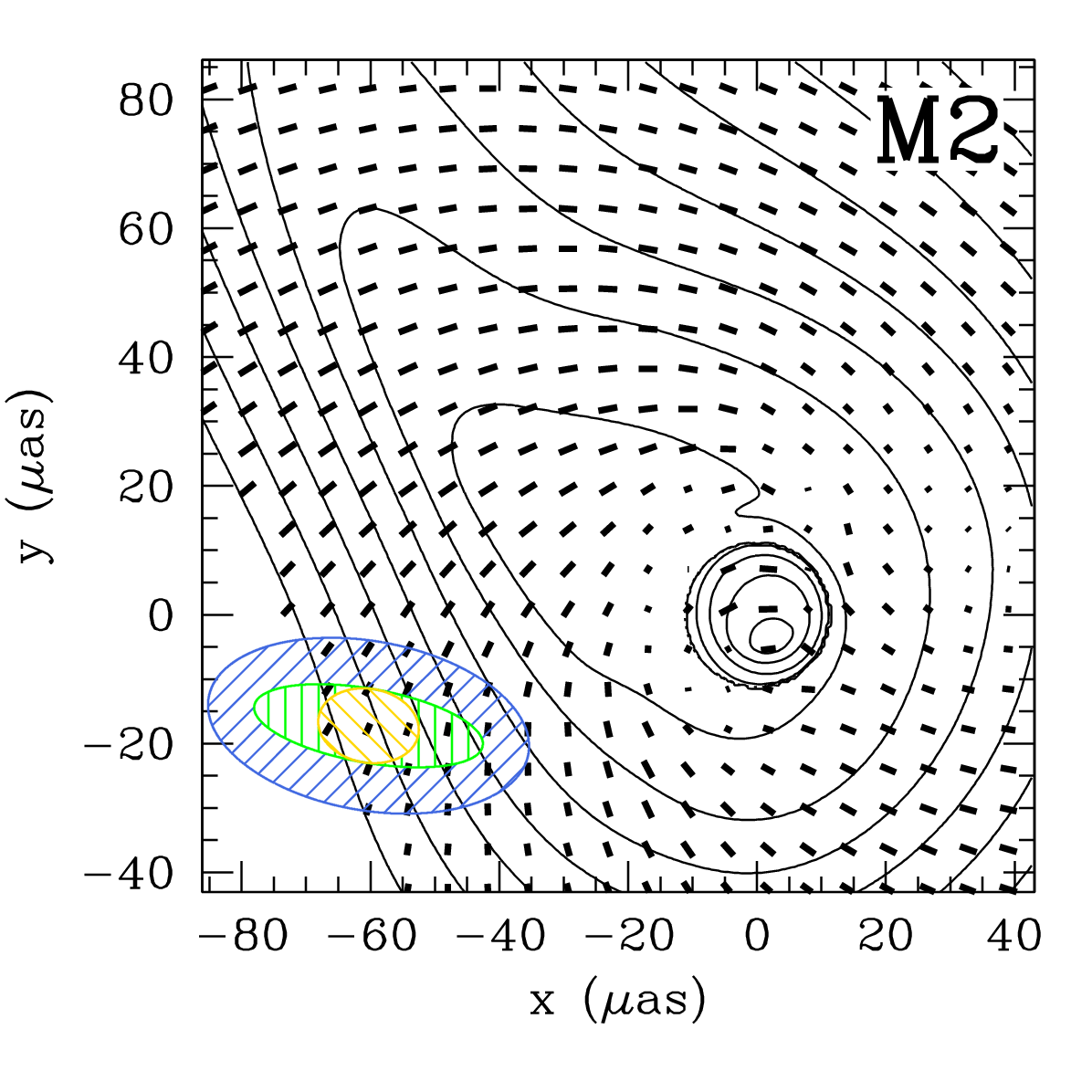}\\
\includegraphics[width=0.3\textwidth]{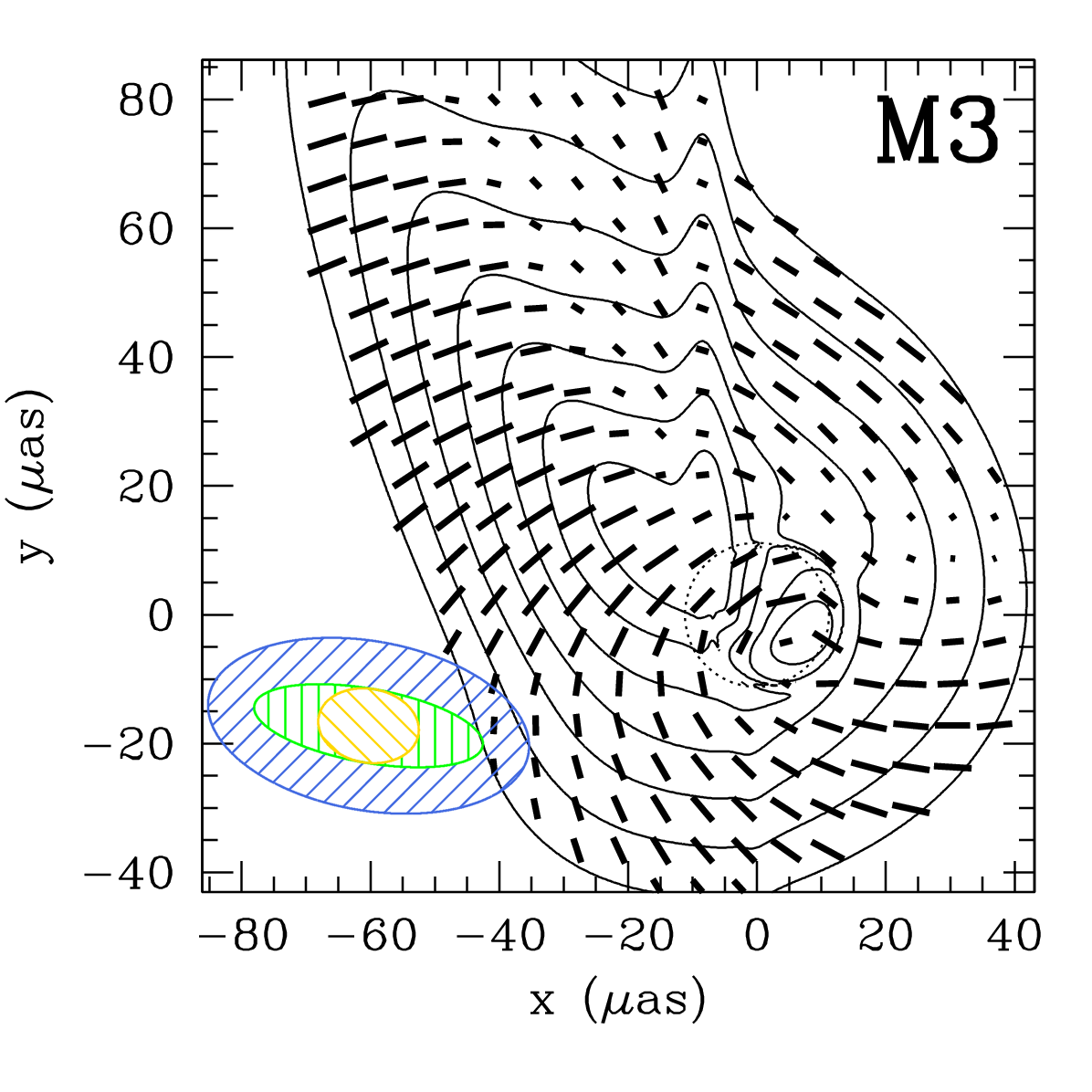}
\includegraphics[width=0.3\textwidth]{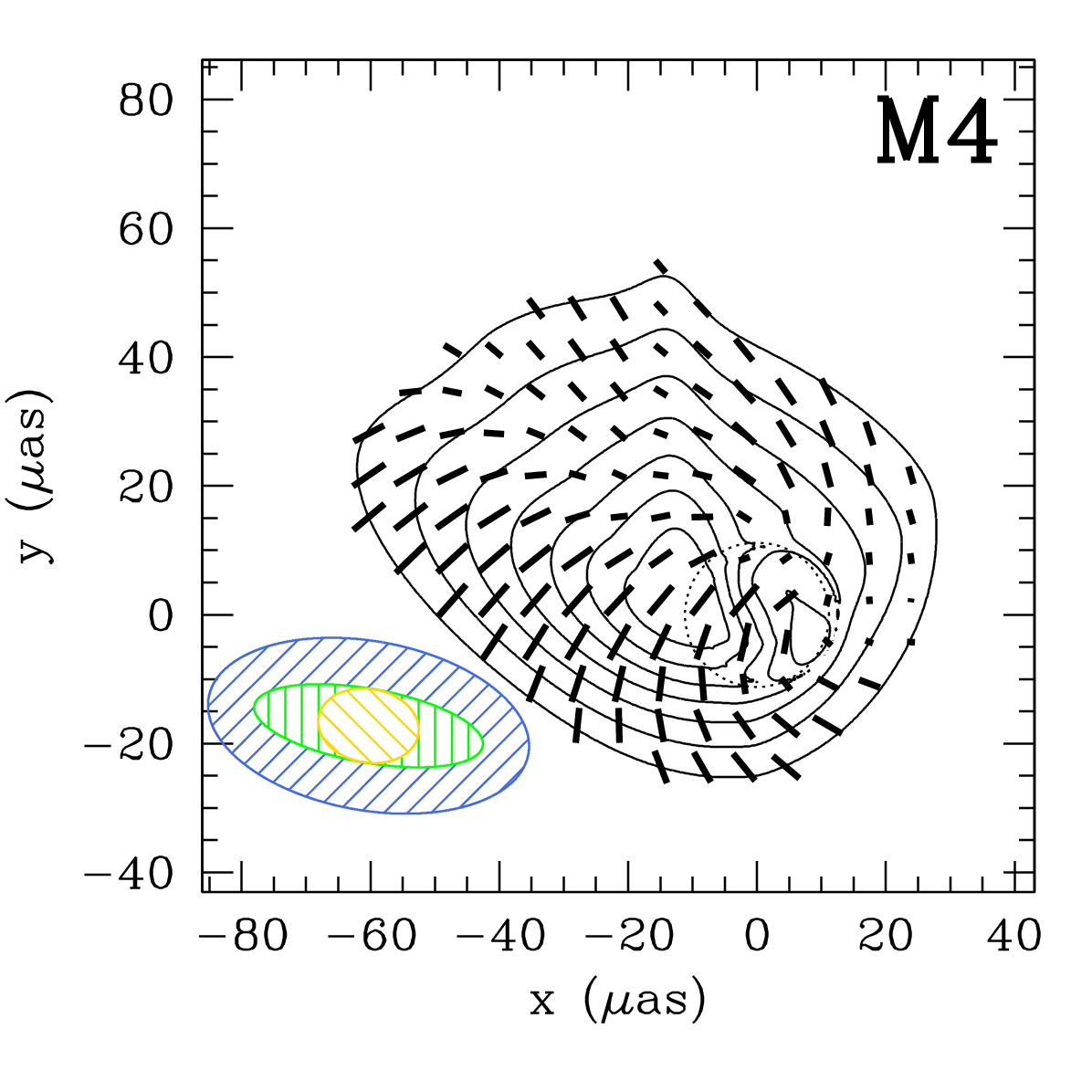}
\includegraphics[width=0.3\textwidth]{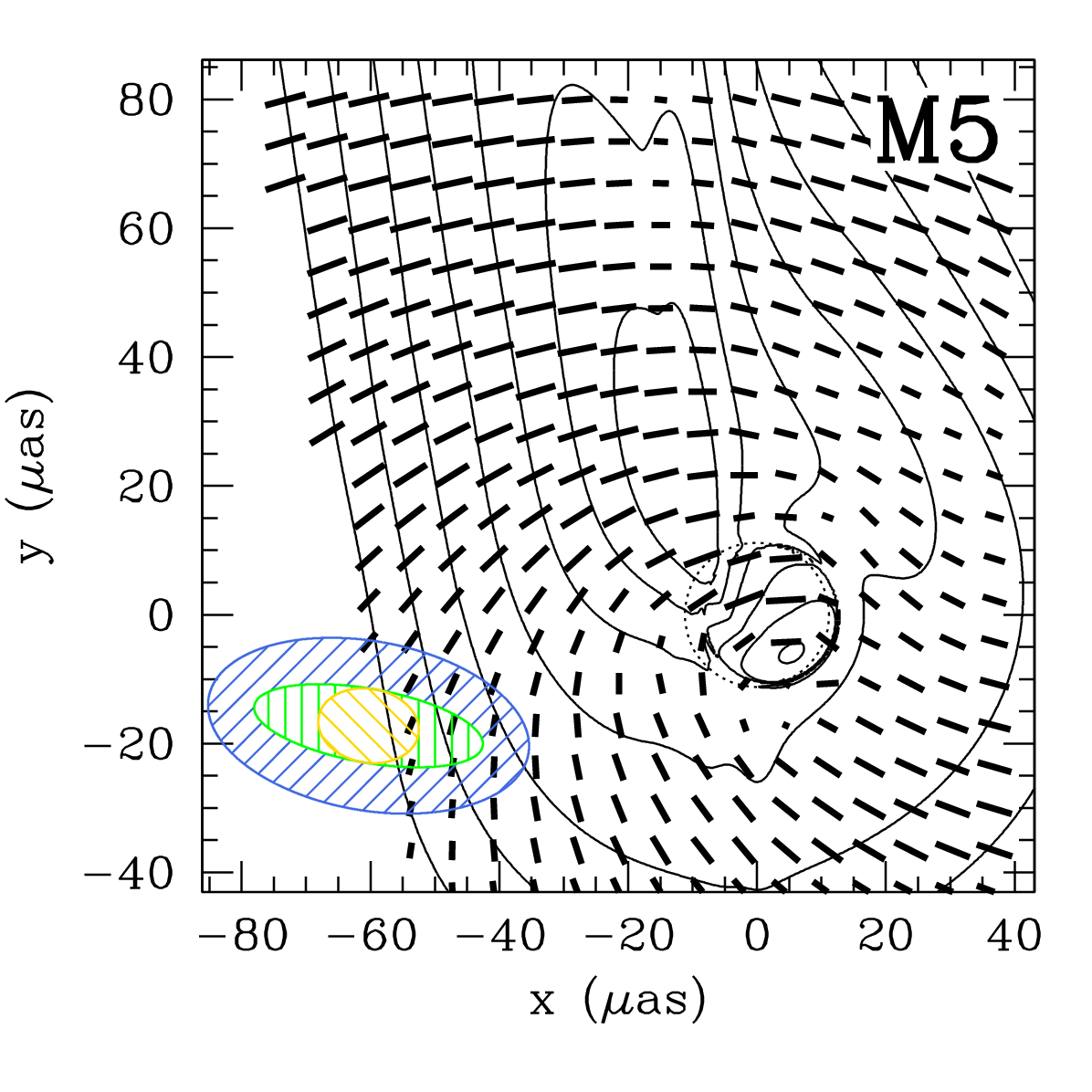}
\end{center}
\caption{$0.87\,\mm$ ($345\,\GHz$) polarization maps of the models in Table
  \ref{tab:models}, superimposed upon logarithmic contours of the
  intensity (factors of 2).  The size of the tick marks are indicative
  of the polarization fraction, with the maximum polarization
  typically being $75\%$.  The polarization vectors have not been
  Faraday rotated.  The ellipses in the lower left
  provide an estimate of the beam size for arrays consisting of the
  North American telescopes only (blue), including the European
  telescopes \& the LMT (green), and all telescopes
  (yellow).\vspace{0.15in}} \label{fig:pol}
\end{figure*}

\subsection{$0.87\,\mm$}

All of the features seen at $1.3\,\mm$ are also present at a
wavelength of $0.87\,\mm$ ($345\,\GHz$).  The two sets of images are subtly
different primarily due to the decreased optical depth.  Despite this,
there are still optically thick regions for all models except M4.
Again it is possible to clearly distinguish footprint size, black hole
spin and collimation model.  At $0.87\,\mm$, the jet orientation is
more easily discerned, though it remains unclear how accurately it may
be determined.

Instead, most important is the reduction in the beam size relative to
the horizon scale.  At submillimeter wavelengths the black hole
silhouette is now larger than the beam, implying that it is possible
to clearly identify it in these images.  However, comparison of the M4
image and the others suggests that it will not be possible to
measure black hole spin with the silhouette detection alone, whose
shape and size is similar for both high and low spins.

This is not unexpected since the primary effect of spin is to shift the
silhouette.  Unlike the horizon, the apparent silhouette size is
insensitve to the spin magnitude.  This can be easily see by
considering critical orbits in the equatorial plane.  The black hole
silhouette is produced by geodesics which terminate on the horizon.
For a non-rotating black hole, these correspond to impact parameters
less than $\sqrt{27} GM/c^2 \simeq 5.2 GM/c^2$, and thus a silhouette
diameter of roughly $10.4\,GM/c^2$.  For a rapidly rotating
black hole, the critical impact parameter for prograde geodesics
decreases.  However, this is compensated for by the increase in the
critical impact parameter of retrograde geodesics.  In the case of a
maximally rotating black hole ($a=1$) these are $2\,GM/c^2$ and
$7\,GM/c^2$, respectively.  Thus the associated silhouette diameter
has shrunk only to $9\,GM/c^2$, though it is shifted by roughly
$2.2\,GM/c^2$ \citep[see, e.g.,][]{Chan:92}.  As a result, efforts to discern the
black hole spin by
careful measurements of the silhouette size and shape alone are likely
to be compromised by the rather large astrophysical uncertainties
associated with the mass and distance to M87's black hole, as well as
the resolution of future experiments.

\section{Jet Polarization} \label{sec:JP}

Polarization maps for the six models we consider are shown in
Fig. \ref{fig:pol}.  The polarized flux can be comparable to the total
flux, with the polarization fraction being largest where the flux is
greatest.  As with earlier radio polarization studies, the
polarization is indicative of the magnetic field structure, and in
particular, the relationship between the plasma density and the
magnetic field.

The maximum polarization fraction in all images is roughly $75\%$, is
strongly correlated with the intensity, and thus should be relatively
easily detected.  The dominant magnetic field orientation is easily
determined from the polarization maps in the usual way: assuming that
the projected magnetic field direction is orthogonal to the
polarization vectors.  Unlike in some quasar jets
\citep{Lyut-Pari-Gabu:05}, the polarization angles here are not
significantly altered by the relativistic motion. This is because of:
{\it (i)} the moderate Lorentz factors in the regions contributing to
the millimeter emission, and {\it (ii)} the orientation of the jet
near the observer's line of sight.  Most importantly, polarization
measurements appear to provide a method by which the structure of the
magnetic field lines may be compared directly with the collimation of
the emitting plasma.

Since in our jet models the emitting electrons are relativistic, it is
not possible to produce a robust estimate of the rotation measure (RM)
from thermal electrons, and thus infer the importance of Faraday
rotation for polarization measurements in M87.  If there is a thermal
population of electrons with a density comparable to the non-thermal
component, we estimate a RM of roughly $10^6\,\rad~\m^{-2}$, which
produces a rotation in the plane of polarization at $1\,\mm$ of
$60^\circ$.  In this case, Faraday depolarization is unlikely to be a
significant problem, and it will be possible to reconstruct the
unrotated polarization map.

This should be tempered, however, by measurements of the rotation
measure within M87's jet at longer wavelengths.  At $20\,\mas$ from
M87's core, the RM has been found to range from
$-4\times10^3\,\rad~\m^{-2}$ to $10^4\,\rad~\m^{-2}$, depending upon
the location observed \citep{Zava-Tayl:02}.  If these are not enhanced
locally, due perhaps to a substantial over density of cold electrons
within the knots being observed, they imply a rotation measure orders
of magnitude higher near the black hole.  That is, if $n\propto
r^{-2}$ and $B\propto r^{-1}$, as expected at large distances, ${\rm
RM}\propto n B r \propto r^{-2}$.  Extrapolating from $20\,\mas$ to
$200\,\muas$ implies a nuclear RM of $10^8\,\rad~\m^{-2}$, two orders
of magnitude larger than our estimate.  In this case,
band-depolarization will still not be important for the largest
bandwidths likely to be accessible (a few GHz).  However, {\em in
situ} Faraday depolarization will play a role, possibly attenuating
the polarized flux considerably, depending upon geometric
considerations.  As such, the absence of polarization itself will
constitute an interesting and potentially important constraint upon
the non-relativistic electron population within M87's outflow.

\section{Conclusions} \label{sec:C}

M87 provides a promising second target for the emerging millimeter and
submillimeter VLBI capability.  Its presence in the Northern sky
simplifies its observation and results in better baseline coverage
than available for Sgr A*.  In addition, its large black-hole mass,
and correspondingly long dynamical timescale, makes possible the use
of Earth aperture synthesis, even during periods of substantial
variability.

Despite being dominated by a jet at millimeter wavelengths, the black
hole silhouette, cast against the base of the counter-jet, is still
clearly present, and may just be resolved at $0.87\,\mm$
($345\,\GHz$).  This requires that both the jet and the counter-jet
are moving slowly near the black hole, and thus relativistic beaming does
not preclude its observation.  If, however, the counter-jet is either
relativistic even at its base or simply not present the silhouette
will not be detectable.  Nonetheless, obtaining a sub-mm image of M87
appears to be more promising than efforts to image the silhouette of
Sgr A* due to the lower opacity of the surrounding synchrotron
emitting gas.  Unfortunately, it is unlikely that the silhouette
alone will be sufficient to constrain the spin without additional
information.  Better constraints may originate from exploring the
effect of spin upon the jet image (via a change in the angular
velocity profile of the jet footprint) or by carefully studying
variability of hot spots in the accretion flow and jetted outflow.
The latter possibility is feasible owing to the low optical depth to
synchrotron self-absorption and the long dynamical timescales in M87.

Direct imaging of M87's jet on horizon scales will necessarily provide
a wealth of information crucial to modern numerical efforts to model
jet formation and propagation.  In particular, such observations will
be sensitive to the jet footprint size, collimation rate, and (if the
jet is truly a magnetically dominated outflow) black hole spin.
Polarized imaging will further shed light upon the structure of the
jet's magnetic field and its relation to the emitting gas.  Less well
constrained will be the jet orientation.

Finally, comparisons of M87 and Sgr A* will provide the first
opportunity to relate the gravitational properties of quasars, about
which much is known, to the lower-mass, dormant black holes that LISA
will access.  Ultimately, such measurements will associate the black
hole demographics measured by LISA with the existing, extensively
studied set of known quasars.

\acknowledgments
We would like to thank Bill Junor, Craig Walker and Chun Ly for
graciously allowing us to use their VLBA images of M87.
This work was supported in part by Harvard University funds.  

\appendix
\section{Force-Free Jet in Covariant Notation}
In order to perform the radiative transfer through the region around
the M87 black hole we must characterize the surrounding plasma.  This
is done in terms of a proper density, $n$, a four-velocity, $u^\mu$,
and the magnetic field within the plasma frame $b^\mu \equiv u_\nu
\mbox{}^*\!F^{\mu\nu}$, where $\mbox{}^*\!F^{\mu\nu}$ is the dual of
the standard electromagnetic field tensor.  Specifying that the
magnetic field be force-free simplifies the definition of these
quantities substantially.  However, usually this is presented in a
3-dimensional vectors instead of the fully covariant form we require.
In this appendix we review some relevant theorems and determine the
jet structure in both notations.

\subsection{Vector Force-Free Theorems}
The ``Force Free'' condition is
\begin{equation}
\rho\E + \j\times\B = 0\,,
\label{eq:vcr}
\end{equation}
i.e., the Lorentz force vanishes.  A direct consequence is that $\E$ and
$\B$ are orthogonal:
\begin{equation}
\B\cdot\left( \rho\E + \j\times\B \right) = \rho\B\cdot\E = 0
\quad\Rightarrow\quad
\E\cdot\B = 0\,.
\label{eq:veb}
\end{equation}
This implies that there exists a frame in which one of these fields
vanishes.  For a force-free solution to exist we require that in this
frame $\E$ vanishes, otherwise there would be no way to satisfy
(\ref{eq:vcr}) generally.  Thus, the Lorentz scalar $B^2-E^2$ must be
non-negative.  The velocity of this frame, called the ``drift
velocity'' is not unique.  However, we can define a class of such
velocities:
\begin{equation}
\v_F = \frac{\E\times\B}{B^2} + \beta\B
\quad\Rightarrow\quad
\E = \v_F\times\B\,.
\label{eq:vvf}
\end{equation}
Choosing $\beta$ determines the frame in which we seek to define the
drift velocity.

Maxwell's equations may be combined with the force-free condition to
produce equations in which only the electromagnetic fields appear.
For stationary solutions these give
\begin{equation}
\div\B = 0\,,\quad
\curl\E = 0\quad\text{and}\quad
\B\times\left(\curl\B\right) = \E\div\E\,,
\label{eq:vff1}
\end{equation}
where the last of these arises directly from the force-free condition
itself.  This may be rewritten in terms of an unknown scalar function,
$\alpha$, as
\begin{equation}
\curl\B = \alpha\B + \v_F\div\E\,,
\label{eq:vff2}
\end{equation}
where $\alpha$ and $\beta \div\E$ are degenerate.  We shall refer to
this second equation as the Force-Free equation.

\subsubsection{Conserved Quantities}
In axisymmetry, $\curl\E=0$ implies that there can be no toroidal
component to $\E$, and thus we may choose $\v_F$ to be purely toroidal
as well.  There are a number of immediate consequences.  Firstly,
\begin{equation}
\B\cdot\grad \alpha
=
\div \alpha\B
=
\div\left( \curl\B - \v_F \div\E \right)
=
0\,,
\label{eq:vca}
\end{equation}
and $\alpha$ is constant along poloidal magnetic field surfaces.
Secondly, inspecting the poloidal components of the Force-Free
equation (in spherical coordinates for convenience):
\begin{equation}
\sqrt{h} B^r = \frac{1}{\alpha} \partial_\theta B_\phi
\quad\text{and}\quad
\sqrt{h} B^\theta = - \frac{1}{\alpha} \partial_r B_\phi\,,
\label{eq:vcBp}
\end{equation}
where $h$ is the determinant of the 3-metric, thus $B_\phi$ is
constant along poloidal field surfaces as well.  Coupled with the
previous result for $\alpha$, this implies that there exists a ``stream
function'', $\psi \equiv \int \alpha^{-1} \d B_\phi$, which satisfies:
\begin{equation}
\sqrt{h} B^r = \partial_\theta \psi
\quad\text{and}\quad
\sqrt{h} B^\theta = -\partial_r \psi\,.
\label{eq:vsf}
\end{equation}
Then, we can define $B_\phi(\psi)$ by its value on the equatorial
plane (or any other convenient surface crossing every poloidal field line),
which implicitly defines $\alpha(\psi) = \d\psi/\d B_\phi$.  Thirdly, defining
$\Omega\equiv v_F^\phi$, this implies:
\begin{equation}
\E = \Omega \ph\times\B
\quad\rightarrow\quad
\curl\E = \ph \B\cdot\grad\Omega = 0\,,
\label{eq:vco}
\end{equation}
and thus $\Omega$ is constant along poloidal field surfaces as well.
Note that for this choice of $\v_F$, there can be light-cylinders, a
region where $v_F>c$.  This does not mean that the Force Free
condition has broken down.  Rather, as we shall explicitly see in the
jet models discussed below, the choice of $\v_F$ may simply be
inappropriate.  Nevertheless, the relationships between $\E$ and $\B$
are unchanged, and we may construct solutions that are continuous in
the relevant observable quantities.

\subsubsection{Determining the Stream Function}
The stream function itself is determined from the phi-component of the
Force-Free equation:
\begin{equation}
\bigg(1-v^\phi v_\phi\bigg) \left(
\partial_r \frac{g_{\theta\theta}}{\sqrt{h}} \partial_r \psi
+
\partial_\theta \frac{g_{rr}}{\sqrt{h}} \partial_\theta \psi
\right)
- 
\frac{v^\phi}{\sqrt{h}}\Bigg[
g_{\theta\theta} \left(\partial_r v_\phi\right)\Big(\partial_r \psi\Big)
+
g_{rr} \left(\partial_\theta v_\phi\right)\Big(\partial_\theta \psi\Big)
\Bigg] + \sqrt{h} B^\phi \frac{\d B_\phi}{\d \psi} = 0\,.
\label{eq:vsff}
\end{equation}
If we recast this relation in cylindrical coordinates, $(R,z,\phi)$, it
reduces to
\begin{equation}
\left(1-R^2\Omega^2\right) \partial_R^2 \psi
+
\left(1-R^2\Omega^2\right) \partial_z^2 \psi
-
\left(1+R^2\Omega^2\right) \frac{1}{R} \partial_R \psi
-
\Omega R^2 \frac{\d \Omega}{\d \psi} \left[
\left(\partial_R \psi\right)^2
+
\left(\partial_z \psi\right)^2
\right]
+
B_\phi \frac{\d B_\phi}{\d \psi}
=
0\,,
\end{equation}
which, after trivial renamings, is identical to Eq. (13) in
\citet{Nara-McKi-Farm:07} \citep[and also][]{Okam:74}.  All that remains
is to specify, $\Omega(\psi)$, $B_\phi(\psi)$ and solve for $\psi$.

\subsection{Vector Jet Structure}

Here we use the various insights gained in the previous section to
write down a qualitative jet model in covariant form.  It is important
to note that in both cases we are attempting only a qualitatively
correct ``solution'', paying specific attention only to ensuring that
the magnetic field geometry and the jet velocity do not become
unphysical anywhere.

Following \citet{Tche-McKi-Nara:08}, we begin by positing a class of
stream functions based upon solutions to the $\Omega=0$
stream-function equation:
\begin{equation}
\psi = r^p \left(1 - \cos \theta \right)\,.
\end{equation}
From this, we may immediately construct the poloidal magnetic field
structure:
\begin{equation}
B^r = r^{p-2}
\quad\text{and}\quad
B^\theta
=
- p r^{p-3}\frac{1-\cos\theta}{\sin\theta}
=
- p r^{p-3} \tan\frac{\theta}{2}
\,.
\end{equation}
Since the toroidal component of the magnetic field scales as $r^{-1}$,
far from the jet source it becomes electromagnetically dominated.
Thus, $B^\phi B_\phi \simeq E^2 = R^2 \Omega^2 B^r
B_r$, and $B_\phi = -R^2 \Omega B^r$, where the negative sign
arises from the fact that the field lines are swept back (recall that
$R=r\sin\theta$ is the cylindrical radius).  That is,
\begin{equation}
B_\phi \simeq - r^2\sin^2\theta \, \Omega \, r^{p-2} \simeq -2 \Omega \psi\,.
\end{equation}
Therefore, since $B_\phi$ is conserved along magnetic field lines,
\begin{equation}
B^\phi
=
\frac{B_\phi(\psi)}{R^2}
=
-\frac{2\Omega(\psi) \psi}{r^2\sin^2\theta}
=
-2\Omega(\psi) r^{p-2} \frac{\tan(\theta/2)}{\sin\theta}
\,.
\end{equation}
Thus, given $\Omega(\psi)$ (which we shall determine below), we now
have the magnetic field value $\B$ everywhere.

\subsubsection{Plasma Velocity}
Because the plasma is non-relativistic initially, and the Bernoulli
constant is conserved along the plasma flow, the plasma velocity is
very similar to the drift velocity in the lab frame, and thus it is
sufficient to compute the latter.  The electric field in the lab frame
is simply $\E = \Omega \ph \times \B$.  Therefore,
\begin{equation}
\v
=
\frac{\E\times\B}{B^2}
=
- \Omega \ph +  \frac{\Omega B_\phi}{B^2} \B 
\end{equation}

\subsubsection{Plasma Density}
Finally, we must determine the electron/pair density in the jet.  We
do this assuming that the continuity equation holds, i.e., $\div n \v
= 0$.  This is a dubious assumption for two reasons.  First, if the
mass-loading of the jet occurs over an extended region (e.g., if it is
due to photon annihilation), then continuity fails locally.  Second,
and more severe, is that we are interested only in the non-thermal tail
of the electron population.  Thus, we might expect the electron
acceleration process and cooling to leave an imprint upon the
non-thermal electron distribution (which is completely ignored in our
prescription).  Nevertheless, in the absence of a more detailed
understanding of jet formation, we proceed with this flawed working
assumption, which gives,
\begin{equation}
\div n \Omega\phi  + \div n\frac{\Omega B_\phi}{B^2} \B
=
\B\cdot\grad n\frac{\Omega B_\phi}{B^2} = 0\,,
\end{equation}
or $n\Omega B_\phi/B^2$ is constant along field lines.  Since $\Omega$
and $B_\phi$ are already known to be constant along field lines, this
gives $n \propto B^2$ along the field line.  Thus, we must only
specify $n/B^2$, i.e., the squared Alfv\'en velocity, at the jet footprint, finding
\begin{equation}
n = B^2 F(\psi)\,,
\end{equation}
everywhere else, for some function $F(\psi)$ that depends upon the jet
loading at the footprint.

\subsection{Covariant Force-Free Theorems}
Each of the results in the previous sections can be presented in a
fully covariant notation as well.  In terms of the electromagnetic
field tensor, the ``Force Free'' condition is
\begin{equation}
F^{\mu\nu}j_\nu = 0\,.
\label{eq:ccr}
\end{equation}
This also directly implies an orthogonality relation:
\begin{equation}
\Fs_{\sigma\mu} F^{\mu\nu} j_\nu
=
-\frac{1}{4} \Fs_{\mu\nu} F^{\mu\nu} j_\sigma
=
0
\quad\Rightarrow\quad
\Fs^{\mu\nu} F_{\mu\nu} = 0\,,
\label{eq:ceb}
\end{equation}
where we used the general relation
${}^*\!A^{\mu\nu}A_{\sigma\nu} = {}^*\!A^{\alpha\nu} A_{\alpha\nu} \delta^\mu_\sigma/4$
for all antisymmetric 2nd-rank tensors, $A^{\mu\nu}$ (which may be
derived directly from the definition of the dual).  This is not
particularly surprising given that the Lorentz invariant $\Fs^{\mu\nu}
F_{\mu\nu} = 2 \E\cdot\B$, where $\E$ and $\B$ are determined in the
locally flat coordinate patch.

Again, we require that it is possible to transform into a frame in
which the electric field vanishes, which corresponds to the condition
that $F^{\mu\nu} F_{\mu\nu} \ge 0$.  As before, the velocity of this
frame is not uniquely defined, nor must it be physical.  However, if
one such velocity is $u_F^\mu$, then we require that the 4-vector that
is coincident with $\E$ in this frame vanishes, i.e.,
\begin{equation}
e^\mu \propto u_F^\nu F^\mu_{~\nu} = 0\,.
\end{equation}
Thus, the acceptable $u_F^\mu$ comprise the null-space of
$F^\mu_{~\nu}$.  Generally, given any $u^\mu$ we can write
$F^{\mu\nu}$ and $\Fs^{\mu\nu}$ as
\begin{equation}
F^{\mu\nu}
=
u^\mu e^\nu - u^\nu e^\mu + \varepsilon^{\mu\nu\alpha\beta}u_\alpha b_\beta
\quad\text{and}\quad
\Fs^{\mu\nu}
=
u^\mu b^\nu - u^\nu b^\mu + \varepsilon^{\mu\nu\alpha\beta}u_\alpha e_\beta\,,
\end{equation}
where $\varepsilon^{\mu\nu\alpha\beta}$ is the Levi-Civita
pseudotensor.  If we set $u^\mu=u_F^\mu$, and thus $e^\mu=0$, this
implies that we can write these as
\begin{equation}
F^{\mu\nu}
=
\varepsilon^{\mu\nu\alpha\beta}{u_F}_\alpha {b_F}_\beta
\quad\text{and}\quad
\Fs^{\mu\nu}
=
u_F^\mu b_F^\nu - u_F^\nu b_F^\mu\,.
\end{equation}
Since we require that there {\em exists} a physical frame in which
$e^\mu$ vanishes, there must be time-like choices for $u_F^\mu$.  This
implies that the null space of $F^{\m\nu}$ is two-dimensional with one
time-like dimension and one space-like dimension (since $u_F^\mu b_\mu
= 0$).

Since we know that $\Fs^\sigma_{~\nu} F^\mu_{~\sigma} = 0$ from the
aforementioned orthogonality condition, the 4-vectors $v^\mu_{(\nu)}
\equiv \Fs^\mu_{~\nu}$ are necessarily components of $F^{\mu\nu}$'s
null space.  In fact, we can show that this space is completely
spanned by the $v^\mu_{(\nu)}$, which are themselves degenerate.  This
may be proved by inspection, constructing a space-like and time-like
basis vectors: $b_F^\nu \Fs^\mu_{~\nu} = b_F^2 u_F^\mu$ and $u_F^\nu
\Fs^\mu_{~\nu} = u_F^2 b_F^\mu$\footnote{Note that
should we choose a form for $u_F^\mu$ that becomes space-like in some
region, $b_F^\mu$ necessarily becomes time-like.}.

Regardless of the nature of $u_F^\mu$, we may always construct
time-like drift velocities.  For any time-like vector, $\eta^\mu$,
\begin{equation}
u^\mu
=
\frac{\Fs^\mu_{~\nu}\Fs^\nu_{~\sigma} \eta^\sigma}
{\sqrt{- \eta_\gamma\Fs_\beta^{~\gamma}\Fs_{\alpha}^{~\beta} \Fs^\alpha_{~\nu}\Fs^\nu_{~\sigma} \eta^\sigma}}
=
\gamma \left(u_F^\mu + \beta b_F^\mu\right)
\quad\text{where}\quad
\beta\equiv\frac{u_F^2 \,b_F^\mu \eta_\mu}{b_F^2 \,u_F^\nu \eta_\nu}
\quad\text{and}\quad
\gamma\equiv -\frac{u_F^2}{\sqrt{-\left(u_F^2+\beta^2 b_F^2\right)}}\,.
\label{eq:cvf}
\end{equation}
Since the quantity $u_F^2$ appears in a number of places, let us
define $\sigma=u_F^2=\pm 1$.  That $\eta^\mu$ be time-like itself is
sufficient to guarantee that $u^\mu$ will be time-like as well.  To
see this, let us decompose $\eta^\mu$ into a tetrad basis with vectors
$e_{(0)}^\mu = u_F^\mu$, $e_{(1)}^\mu \propto b_F^\mu$ and
$e_{(2,3)}^\mu$ orthogonal to the rest (and thus 
necessarily space-like).  Then,
\begin{equation}
u^2 \propto {\eta^{(0)}}^2 e_{(0)}^2 + {\eta^{(1)}}^2 e_{(1)}^2
\le
{\eta^{(0)}}^2 e_{(0)}^2 + {\eta^{(1)}}^2 e_{(1)}^2 + {\eta^{(3)}}^2 e_{(3)}^2 + {\eta^{(3)}}^2 e_{(3)}^2
=
\eta^2
< 0\,.
\end{equation}
Finally, we have chosen the arbitrary sign of $u^\mu$ such that it
corresponds to forward-propagating particles.  To see this generally,
we again appeal to the tetrad expansion of $\eta$:
\begin{equation}
u^\mu \eta_\mu
=
\gamma\left( u_F^\mu \eta_\mu + \beta b_F^\mu \eta_\mu \right)
=
\frac{\gamma u_F^2}{u_F^\nu\eta_\nu}
\left[
\frac{\left( u_F^\mu \eta_\mu \right)^2}{u_F^2}
+ 
\frac{\left( b_F^\mu \eta_\mu \right)^2}{b_F^2}
\right]
=
\frac{\gamma \sigma}{u_F^\nu \eta_\nu}
\left( {\eta^{(0)}}^2 e_{(0)}^2 + {\eta^{(1)}}^2 e_{(1)}^2 \right)\,.
\end{equation}
Since the term in the parentheses is negative and we choose $u_F^\nu \eta_\nu$
to be negative by convention, this implies that $\gamma$ must have the
same sign as $\sigma$.

Note that if $\eta^\mu \propto u_F^\mu$, we find $u^\mu = u_F^\mu$.  Thus, 
choosing $\eta^\mu$ is equivalent to choosing the frame in which we
wish to define the drift velocity.  That these are not all equivalent
is a consequence of the ambiguity under boosts along the magnetic
field.  Nevertheless, if we are given a preferred frame (e.g., the lab
frame) there is a unique definition, as in the 3-D vector
computation. By inspection in the locally-flat coordinate frame we can
show that this definition of $u^\mu$ is identical to the 3-D vector
case.

We now turn to the various representations of the Force-Free
equation.  Recalling that Maxwell's equations give us
$\nabla_\nu F^{\mu\nu} = 4\pi j^\mu$, the force-free condition may be
trivially written in terms of $F^{\mu\nu}$:
\begin{equation}
F^\mu_{~\nu} \nabla_\sigma F^{\nu\sigma} = 0\,,
\quad\text{or}\quad
\nabla_\sigma F^{\mu\sigma}
=
\alpha b_F^\mu
+
\zeta u_F^\mu\,,
\label{eq:cff}
\end{equation}
i.e., once again we find that the vector we desire (in this case
$j^\mu$) is in the null-space of $F^\mu_{~\nu}$.  Now, there are two
undefined functions, $\alpha$ which is analogous to the 3-D vector
case, and $\zeta$.  As it turns out, we will be able to completely
determine the magnetic field without ever having to specify $\zeta$,
which is not surprising given that there are only three independent
components to $b^\mu$.

\subsubsection{Conserved Quantities}
First, let us derive a number of useful relations in axisymmetry.  As
before, we shall assume that it is possible to put the field-line
velocity (drift velocity) into a purely toroidal form by an
appropriate boost along $b_F^\mu$:
\begin{equation}
u_F^\mu = u_F^t t^\mu + u_F^\phi \phi^\mu\,,
\end{equation}
where $t^\mu$ and $\phi^\mu$ are the time-like and azimuthal Killing
vectors present in all stationary, axisymmetric spacetimes.  The first
thing we note is:
\begin{equation}
0
=
{b_F}_\mu \nabla_\nu \Fs^{\mu\nu}
=
{b_F}_\mu \nabla_\nu \left( u_F^\mu b_F^\nu - b_F^\mu u_F^\nu \right)
=
 b_F^\mu b_F^\nu \nabla_\nu {u_F}_\mu\,,
\end{equation}
and thus the magnetic field must lie upon non-shearing velocity
surfaces.  This may be simplified by Killing's equation and the
condition that
${b_F}_\mu u_F^\mu = 0 \rightarrow {b_F}_t = -{b_F}_\phi \Omega$
where $\Omega\equiv u_F^\phi/u_F^t$:
\begin{equation}
b_F^\mu b_F^\nu \nabla_\nu {u_F}_\mu
=
b_F^\mu b_F^\nu \left( t_\mu \partial_\nu u_F^t + \phi_\mu \partial_\nu u_F^\phi \right)
=
{b_F}_\phi b_F^\nu \left( \partial_\nu \Omega u_F^t - \Omega \partial_\nu u_F^t \right)
=
{b_F}_\phi u_F^t b_F^\nu \partial_\nu \Omega\,.
\label{eq:cco}
\end{equation}
That is, $\Omega$ is constant along magnetic field lines.

Next we note
\begin{equation}
0
=
{u_F}_\mu \nabla_\nu \Fs^{\mu\nu}
=
{u_F}_\mu \nabla_\nu \left( u_F^\mu b_F^\nu - b_F^\mu u_F^\nu \right)
=
b_F^\mu u_F^\nu \nabla_\nu {u_F}_\mu + \sigma \nabla_\nu b_F^\nu\,.
\end{equation}
The first term may again be simplified using Killing's equation:
\begin{equation}
\begin{aligned}
b_F^\mu u_F^\nu \nabla_\nu {u_F}_\mu
&=
b_F^\mu u_F^\nu \nabla_\nu \left( u_F^t t_\mu + u_F^\phi \phi_\mu \right)\\
&=
b_F^\mu \left(
t_\mu u_F^\nu\partial_\nu u_F^t + \phi_\mu u_F^\nu\partial_\nu u_F^\phi
+
u_F^t u_F^\nu\nabla_\nu t_\mu + u_F^\phi u_F^\nu\nabla_\nu \phi_\mu
\right)\\
&=
-b_F^\mu \left(
u_F^t u_F^\nu\nabla_\mu t_\nu + u_F^\phi u_F^\nu\nabla_\mu \phi_\nu
\right)\\
&=
b_F^\mu \left(
t_\nu u_F^\nu\partial_\mu u_F^t + \phi_\nu u_F^\nu\partial_\mu u_F^\phi
-
u_F^\nu \nabla_\mu {u_F}_\nu
\right)\\
&=
b_F^\mu \left( {u_F}_t \partial_\mu u_F^t + {u_F}_\phi \partial_\mu u_F^\phi \right)\\
&=
b_F^\mu \left( {u_F}_t \partial_\mu u_F^t + {u_F}_\phi \partial_\mu \Omega u_F^t \right)\\
&=
b_F^\mu \left[ \left({u_F}_t + {u_F}_\phi \Omega\right)\partial_\mu u_F^t
+ {u_F}_\phi u_F^t \partial_\mu \Omega \right]\\
&=
\sigma b_F^\mu \partial_\mu \ln u_F^t\,.
\end{aligned}
\end{equation}
Therefore, $\nabla_\nu b_F^\nu = - b_F^\mu \partial_\mu \ln u_F^t$.
However, from the Force-Free equation, we have
\begin{equation}
0
=
\nabla_\mu \nabla_\nu F^{\mu\nu}
=
\nabla_\mu \alpha b_F^\mu + \nabla_\mu \zeta u_F^\mu
=
b_F^\mu \partial_\mu \alpha + \alpha \nabla_\mu b_F^\mu
=
b_F^\mu \partial_\mu \alpha - \alpha b_F^\mu \partial_\mu \ln u_F^t
=
u_F^t b_F^\mu \partial_\mu \frac{\alpha}{u_F^t}\,,
\label{eq:cca}
\end{equation}
and thus $\alpha/u_F^t$ is conserved along field lines.

This now brings us to the stream function.  Again, looking at the
poloidal parts of the Force-Free equation we find:
\begin{equation}
\begin{aligned}
\alpha b^r
&=
\varepsilon^{r\nu\alpha\beta} \partial_\nu {u_F}_\alpha {b_F}_\beta
=
\frac{1}{\sqrt{-g}}
\partial_\theta \left( {u_F}_\phi {b_F}_t - {u_F}_t {b_F}_\phi \right)
=
-\frac{\sigma}{\sqrt{-g}}
\partial_\theta \left(\frac{{b_F}_\phi}{u_F^t}\right)\\
\alpha b^\theta
&=
\varepsilon^{\theta\nu\alpha\beta} \partial_\nu {u_F}_\alpha {b_F}_\beta
=
-\frac{1}{\sqrt{-g}}
\partial_r \left( {u_F}_\phi {b_F}_t - {u_F}_t {b_F}_\phi \right)
=
\frac{\sigma}{\sqrt{-g}}
\partial_r \left(\frac{{b_F}_\phi}{u_F^t}\right)\,,\\
\end{aligned}
\end{equation}
where we used 
\begin{equation}
{u_F}_t {b_F}_\phi - {u_F}_\phi {b_F}_t
= 
{u_F}_t {b_F}_\phi + {u_F}_\phi {b_F}_\phi \frac{u_F^\phi}{u_F^t}
=
\frac{{b_F}_\phi}{u^t} \left( {u_F}_t u_F^t + {u_F}_\phi u_F^\phi
\right)
=
\sigma \frac{{b_F}_\phi}{u_F^t}\,.
\end{equation}
Since $\alpha/u^t$ is conserved along field lines, we may define a
stream function by
\begin{equation}
\psi = \int \frac{u_F^t}{\alpha} \,\d\!\left(\frac{{b_F}_\phi}{u_F^t}\right)\,,
\end{equation}
which is very similar to the 3-D vector case, except for the presence
of the $u_F^t$.  In terms of this, we have
\begin{equation}
\frac{\sqrt{-g}}{u_F^t} b_F^r = - \sigma \partial_\theta\psi
\quad\text{and}\quad
\frac{\sqrt{-g}}{u_F^t} b_F^\theta = \sigma \partial_r\psi\,.
\label{eq:csf}
\end{equation}
Now the problem is again reduced to determining $\psi$.

\subsubsection{Determining the Stream Function}
At first glance, it may appear that we must also know $\zeta$ to
obtain an equation for $\psi$.  However, we can use the fact that
$b_F^\mu {u_F}_\mu = 0$ to avoid this requirement:
\begin{equation}
\begin{aligned}
\sqrt{-g} \alpha b_F^2
&=
\sqrt{-g} {b_F}_\mu \nabla_\nu F^{\mu\nu}
=
\sqrt{-g} \epsilon^{\mu\nu\alpha\beta}{b_F}_\mu \partial_\nu {u_F}_\alpha {b_F}_\beta
=
\sqrt{-g} \epsilon^{\mu\nu\alpha\beta}{b_F}_\mu {u_F}_\alpha \partial_\nu {b_F}_\beta\\
&=
-\left( {b_F}_t {u_F}_\phi - {b_F}_\phi {u_F}_t \right)
\Big( \partial_r {b_F}_\theta - \partial_\theta {b_F}_r \Big)
+ {b_F}_\theta \partial_r \left(\frac{{b_F}_\phi}{u_F^t}\right)
- {b_F}_r \partial_\theta \left(\frac{{b_F}_\phi}{u_F^t}\right)\\
&\qquad\qquad
- {b_F}_\theta \left( {b_F}_t \partial_r {u_F}_\phi - {b_F}_\phi \partial_r {u_F}_t \right)
+ {b_F}_r \left( {b_F}_t \partial_\theta {u_F}_\phi - {b_F}_\phi \partial_\theta {u_F}_t \right)\\
&=
\sigma \sqrt{-g} \alpha \Big( {b_F}_r b_F^r + {b_F}_\theta b_F^\theta \Big)
+ 
\sigma \frac{{b_F}_\phi}{u_F^t} \Bigg[
\partial_r \frac{g_{\theta\theta}}{u_F^t \sqrt{-g}} \partial_r \psi
+
\partial_\theta \frac{g_{rr}}{u_F^t \sqrt{-g}} \partial_\theta \psi\\
&\qquad\qquad
- {b_F}_\theta \left( u_F^\phi \partial_r {u_F}_\phi + u_F^t \partial_r {u_F}_t \right)
+ {b_F}_r \left( u_F^\phi \partial_\theta {u_F}_\phi + u_F^t \partial_\theta {u_F}_t \right)
\Bigg]\,,
\end{aligned}
\end{equation}
where the following relations were used,
\begin{equation}
\begin{gathered}
{b_F}_t \partial_i {u_F}_\phi - {b_F}_\phi \partial_i {u_F}_t
=
-\Omega {b_F}_\phi \partial_i {u_F}_\phi - {b_F}_\phi \partial_i {u_F}_t
=
-\frac{{b_F}_\phi}{u_F^t} \left(
u_F^\phi \partial_i  {u_F}_\phi + u_F^t \partial_i {u_F}_t
\right)\\
{b_F}_\theta \partial_r \left(\frac{{b_F}_\phi}{u_F^t}\right)
- {b_F}_r \partial_\theta \left(\frac{{b_F}_\phi}{u_F^t}\right)
=
\sigma \sqrt{-g}\alpha \left( {b_F}_r b_F^r + {b_F}_\theta b_F^\theta \right)\,.
\end{gathered}
\end{equation}
Note also that
\begin{equation}
b_F^2
=
{b_F}_t b_F^t + {b_F}_\phi b_F^\phi
+
{b_F}_r b_F^r + {b_F}_\theta b_F^\theta
=
{b_F}_r b_F^r + {b_F}_\theta b_F^\theta
+
\frac{{b_F}_\phi}{u_F^t} \left( u_F^t b_F^\phi - u_F^\phi b_F^t \right)\,,
\end{equation}
and thus,
\begin{multline}
\partial_r \frac{g_{\theta\theta}}{u_F^t \sqrt{-g}} \partial_r \psi
+
\partial_\theta \frac{g_{rr}}{u_F^t \sqrt{-g}} \partial_\theta \psi
-
\sigma {b_F}_\theta \left( u_F^\phi \partial_r {u_F}_\phi + u_F^t \partial_r {u_F}_t \right)\\
+
\sigma {b_F}_r \left( u_F^\phi \partial_\theta {u_F}_\phi + u_F^t \partial_\theta {u_F}_t \right)
+
\sqrt{-g} \frac{\d \left({b_F}_\phi/u_F^t\right)}{\d \psi}
\left( \frac{b_F^\phi}{{u_F}_t} \right) = 0\,.
\label{eq:csff}
\end{multline}
This is quite similar to Eq. (\ref{eq:vsff}), but significant
differences remain.  Most notably are the presence of various
flow-velocity components (the $u_F^t$'s).  When $\Omega=0$ this
simplifies somewhat, though not as completely as in the 3-D vector
case.  The reason is that there may still be non-negligible terms due
to gravity (i.e., non-vanishing derivatives of $g_{tt}$).  In flat
space, however, this gives precisely the same expression as
Eq. (\ref{eq:vsff}).

\subsection{Covariant Jet Structure}
Again, let us begin with the the class of stream functions,
\begin{equation}
\psi = r^p \left( 1- \cos\theta \right)\,.
\end{equation}
Furthermore, at each location we have $u^t$ in terms of $x^\mu$ and
$\Omega(\psi)$:
\begin{equation}
u_{F}^t = \left| g_{tt} + 2 g_{t\phi}\Omega + g_{\phi\phi}\Omega^2 \right|^{-1/2}\,.
\end{equation}
where some care has been taken to deal with the possibility of
$u_F^\mu$ being space-like.
We then determine the poloidal magnetic field seen by the orbiting
observer as
\begin{equation}
\begin{gathered}
b_{F}^r = -\sigma \frac{r^p\sin\theta}{u_{F}^t \sqrt{-g}}
\,,\quad
b_{F}^\theta = -\sigma p \frac{ r^{p-1} (1-\cos\theta)}{u_{F}^t \sqrt{-g}}\\
{b_{F}}_\phi = -2\Omega\psi \, u_{F}^t
\,,\quad
{b_{F}}_t = - {b_{F}}_\phi \Omega
\end{gathered}
\end{equation}
All of these functions are well behaved in the region of interest.
Note, however, that until we settle upon the plasma velocity, these
results do {\em not} provide the magnetic field in the jet frame.

\subsubsection{Plasma Velocity}
Again, we assume that the plasma velocity is similar to the drift
velocity as measured in some lab frame.  Based upon the physical
notion that zero-angular momentum gas should not see an acceleration
along the jet (and guided by the requirement that we have a meaningful
time-like vector), we choose
\begin{equation}
\eta_\mu = \left(-\frac{1}{\sqrt{-g^{tt}}},0,0,0 \right)\,.
\end{equation}
As a result,
\begin{equation}
u^\mu = \gamma \left(u_F^\mu + \beta b_F^\mu\right)
\quad\text{where}\quad
\beta = \frac{\sigma b_F^t}{b_F^2 u_F^t}\,,
\end{equation}
and $\gamma$ is given by Eq. (\ref{eq:cvf}).
In this frame, the magnetic field is
\begin{equation}
b^\mu
=
u_\nu  \Fs^{\mu\nu}
=
-u_\nu \sigma \left(  u_F^\mu b_F^\nu - b_F^\mu u_F^\nu \right)
=
\gamma\left(b_F^\mu - \sigma b_F^2 \beta u_F^\mu\right)\,,
\end{equation}
where the additional factor of $-\sigma$ is necessary to keep
$\Fs^{\mu\nu}$ continuous across the light cylinder\footnote{Up to now
  we did not need to worry about this since the overall sign of
  $\Fs^{\mu\nu}$ did not matter}.

\subsubsection{Plasma Density}
Again, we set the density using the continuity equation.  This gives
\begin{equation}
\nabla_\mu n u^\mu
=
\nabla_\mu n\gamma\left( u_F^\mu + \beta b_F^\mu \right)
=
\gamma \beta n \nabla_\mu b_F^\mu + b_F^\mu \partial_\mu \gamma\beta n
=
u_F^t b_F^\mu \partial_\mu \frac{\gamma\beta n}{u^t}
=
0\,.
\end{equation}
Thus, similar to the 3-D vector case, some combination of the magnetic
field, $\Omega$ and the density is conserved along field
lines.  In Boyer-Lindquist coordinates, we can simplify $\beta$
somewhat to bring this expression more in line with the 3-D vector
case:
\begin{equation}
\beta
=
\frac{\sigma b_F^t}{b_F^2 u_F^t}
=
- \frac{\sigma b_F^\phi {u_F}_\phi}{b_F^2 u_F^t {u_F}_t}
=
\frac{\sigma {b_F}_\phi u_F^\phi}{b_F^2 u_F^t {u_F}_t}
\left(g^{\phi\phi}-\Omega g^{t\phi}\right)
\left(g_{\phi\phi}-\frac{g_{t\phi}}{\Omega}\right)
=
\sigma \left(\frac{{b_F}_\phi}{u_F^t}\right) \Omega
\frac{1}{b_F^2}
\left( g^{tt} + \frac{1}{\Omega} g^{t\phi}\right)\,.
\end{equation}
The first two non-trivial factors, ${b_F}_\phi/u_F^t$ and $\Omega$, are
conserved along field lines, which is very similar to the 3-D vector
case.  Therefore,
\begin{equation}
b_F^\mu \partial_\mu
\left[
\frac{\gamma n}{u_F^t b_F^2}
\left(g^{tt} + \frac{g^{t\phi}}{\Omega}\right)
\right]
=
0\,,
\end{equation}
is conserved along field lines.  That is, the density is given by
\begin{equation}
n = \frac{u_F^t b_F^2}{\gamma} \left( g^{tt} + \frac{g^{t\phi}}{\Omega} \right)^{-1} F(\psi)
\end{equation}
given the arbitrary function $F(\psi)$.  This is chosen as described
in the main text.

\subsubsection{Parameters \& Summary}
As before we must set three parameters (the power-law index $p$ and
the density \& magnetic field normalizations) and two functions of
$\psi$ ($\Omega(\psi)$ and $F(\psi)$).  We describe how these are
chosen in the main text.  Given these, in practice the procedure for
computing the magnetic field, plasma velocity and plasma density at a
point $x^\mu$ is:
\begin{enumerate}
\item Settle upon $p$, $\Omega(\psi)$ and $F(\psi)$.
\item Construct $b_F(x^\mu)$ and $u_F^t(x^\mu)$.
\item From these, construct $b^\mu(x^\mu)$, $u^\mu(x^\mu)$ and $n(x^\mu)$.
\end{enumerate}

\end{document}